\documentclass[english,showpacs,floatfix]{revtex4}

\usepackage[dvips]{graphicx}
\usepackage{longtable}
\usepackage{amsmath,amssymb}
\usepackage{dcolumn}
\usepackage{latexsym}
\usepackage{babel}
\usepackage[latin1]{inputenc}
\begin{document}
\title{The energy absorption problem of a brane-world black hole}
\author{Lihui Liu}
\author{Bin Wang}
\email{wangb@fudan.edu.cn}
\affiliation{Department of Physics, Fudan University, Shanghai
200433, People's Republic of China }

\author{Guohong Yang}
\affiliation{Department of Physics, Shanghai University, People's Republic of China }


\begin{abstract}
We have studied the wave dynamics and the energy absorption
problem for the scalar field as well as the brane-localized
gravitational field in the background of a braneworld black hole.
Comparing our results with the four-dimensional Schwarzschild
black hole, we have observed the signature of the extra dimension
in the energy absorption spectrum.
\end{abstract}

\maketitle

\section{Introduction}
Recent developments on higher dimensional gravity resulted in a
number of interesting theoretical ideas such as the brane-world
concept in which the standard model fields are confined to our
four-dimensional world viewed as a hyperspace embedded in the
higher-dimensional spacetime where gravity can propagate\cite{s1,
randall, s3}. The simplest models in this context are proposed by
Randall and Sundrum\cite{randall}. It was argued that the extra
dimensions need not be compact and in particular it was shown that
it is possible to localize gravity on a 3-brane when there is one
infinite extra dimension \cite{s1, randall}. A striking
consequence of the theory with large extra dimensions is that it
can lower the fundamental gravity scale and allow the production
of mini black holes in the universe. Such mini black holes are
centered on the brane and may have been created in the early
universe due to density perturbations and phase transitions.
Recently it was proposed that such mini black holes may also be
produced in high energy collisions and could be probed in particle
accelerator experiments in the very near future\cite{s4,s5,s6,s7}.

A great deal of effort has been expanded for the precise
determination of observational signatures of such mini black
holes. One among them is the black hole quasinormal modes (QNM),
which can disclose the evolution of the perturbation around black
hole during a certain time interval and was argued carrying unique
fingerprint of the black hole existence and is expected to be
detected through gravitational wave observations in the near
future (see reviews on this topic and references therein
\cite{s8}). Recently the QNMs of a brane-world black hole have
been studied in \cite{eabdalla,abdalla,shen}. The gravitational
perturbation of the mini black hole would be a characteristic
sound and could tell us the existence of such black hole. Another
chief possibility of observing this kind of mini black hole is the
spectrum of Hawking radiation which is expected to be detected in
particle accelerator experiments \cite{s9,s10,s11,s14,s12,park,shen,Horowitz2,park2,kaloper2}.
Since such small black holes carry information of extra dimensions
and persist different properties compared to those of ordinary
4-dimensional black holes, these two tools of detecting mini black
holes can help to read the extra dimensions.

Most of the available works have been done for the idealized case
by studying the brane-localized modes in the QNM and Hawking
radiation. One of the reason is that in the brane world scenario
it was assumed that most standard matter fields are
brane-localized. In the study of Hawking radiation, it was argued
that the emission on the brane is dominated compared to that off
the brane\cite{Horowitz}. This argument was supported numerically
in the case of standard model field emission by the (4 +
n)-dimensional non-rotating black holes \cite{s12,park} and also
in the higher-dimensional rotating black hole
background\cite{Horowitz2}. Some counterexamples to this argument
have also been found\cite{park2}. However in general it is very
hard to obtain exact soltions of higher-dimensional Einstein
equations so that the knowledge of the bulk is lacking, the
thorough study on the field emission in the bulk is still a
challenging task. Using a recently constructed exact black hole
localized on a 3-brane in a world with two large extra dimensions
\cite{kaloper}, Dai et al explored the Hawking decay channels with
the influence of the brane tension\cite{kaloper2}. They found for
the non-rotating black holes the dominate channels are still the
brane-localized modes.

In this paper we are going to study a braneworld black hole
obtained in \cite{casadio}. A general class of spherically
symmetric and static solution to the field equation with a
5-dimensional cosmological constant can be derived by considering
a general line element of the type
\begin{equation}\label{e3}
    ds^2=A(r)dt^2-\frac{1}{B(r)}dr^2-r^2 d\Omega^2,
\end{equation}
and relaxing the condition $A(r)=B^{-1}(r)$ used in obtaining the
usual 4-dimensional black holes. Casadio et al obtained two types
of solutions by fixing either $A(r)$ or $B(r)$ and in one case
with the choice $A(r)=1-2M/r$, the metric reads \cite{casadio}
\begin{equation}\label{e0}
    ds^2=(1-\frac{2M}{r})dt^2-\frac{1-\frac{3M}{2r}}{\left(1-\frac{2M}{r}\right)
  \left(1-\frac{\gamma M}{2r}\right)}dr^2-r^2 d\Omega^2.
\end{equation}
This braneworld black hole is called CFM black hole. It would be
noted that Eq.(2) was not derived from 5-dimensional solution, but
rather was obtained as a solution to 4-dimensional equations
constraining the possible form of the 5-dimensional metric. The
extension of the asymptotically flat static spherically metric on
the brane into the bulk has been discussed in \cite{dd}. The usual
4-dimensional Schwarzschild black hole is recovered with
$\gamma=3$. The corresponding Hawking temperature is given
by\cite{casadio} $T_{BH}=\sqrt{1-3(\gamma-3)/2}/(8\pi M)$. Thus in
comparison with the Schwarzschild black hole, the braneworld black
hole will either be hotter or colder depending upon the sign of
$\gamma-3$. In this work we are restricted to the case when
$\gamma\leq 11/3$ as in \cite{eabdalla} to ensure the Hawking
temperature to be physical. The QNM of this kind of braneworld
black hole was studied in \cite{eabdalla}.  We will concentrate on
the absorption problem of this black hole. Immersed in external
radiation fields, the black hole scatters the radiation and may
either absorb or amplify it. The process is described by dynamical
perturbation equations in the black hole background. We will study
the scalar and axial gravitational perturbations on the background
of this CFM black hole and investigate the absorption problem. We
will extract the information on the extra dimensional influence on
the absorption of the braneworld black hole by comparing to that
of the usual four-dimensional black hole.

For future convenience, we adopt the notation in \cite{Ch},
\begin{equation}\label{e4}
    ds^2=e^{2\nu}dt^2-e^{2\psi}d\varphi^2-e^{2\mu_2}dr^2-e^{2\mu_3}d\theta^2.
\end{equation}
Comparing with (2), we have
\begin{equation}\label{e00}
    e^{2\nu}=A(r),\ \ e^{-2\mu_2}=B(r),\ \ e^{2\mu_3}=r^2,\ \
    e^{2\psi}=r^2\sin^2\theta.
\end{equation}

In the vicinity of the event horizon $r_H$, the asymptotic
expressions of the metric components $A(r)$ and $B(r)$ can be
written in the first order
\begin{eqnarray}\label{e5}
    &&A(r)\sim A_1(r/r_H-1), \nonumber \\
    &&B(r)\sim B_1(r/r_H-1),\ \ (r\rightarrow r_H+0),
\end{eqnarray}
where $A_1=1$ and $B_1=4-\gamma$ are dimensionless, positive real
numbers. Demanding that the spacetime is asymptotically flat, we
have,
\begin{eqnarray}\label{e6}
    &&A(r)\rightarrow 1,\nonumber\\
    &&B(r)\rightarrow 1,\ \ (r\rightarrow \infty).
\end{eqnarray}

Our paper is organized as follows: in \S II we will go over scalar
perturbation and investigate its energy absorption; in \S III we
will derive the wave equation of the gravitational perturbation
and study its energy absorption spectra. We will sumarize our
results in the last section.

\section{ Scalar Perturbation and Its energy  Absorption}

For simplicity we first consider the braneworld
black hole immersed in the massless scalar field confined on the
brane. We will go over the scalar perturbation and study the energy
absorption problem in the general four dimensional spherical
metric, Eq. (\ref{e3}) and then apply the general results to
the CFM black hole.

\subsection{Scalar Perturbation Equations}

The scalar perturbation is governed by the Klein-Gorden equation.
For the massless scalar field, we have
\begin{equation}\label{f1}
    \Box\Phi=(-g)^{-1/2}\partial_{\mu}\left[(-g)^{1/2}g^{\mu\nu}
    \partial_{\nu} \Phi\right]=0.
\end{equation}
Using the decomposition of the scalar field
\begin{equation}\label{f2}
    \Phi=\sum_{l=0}^{\infty}R_l(r)
    P_l(\cos\theta)e^{-i\omega t},
\end{equation}
we have the Schr\"{o}dinger like radial wave equations,
\begin{equation}\label{f3}
    \frac{d^2(rR_l)}{dr_*^2}
    +\left[\omega^2-V_{s,l}(r)\right](rR_l)=0.
\end{equation}
$r_*$ is the tortoise coordinate defined as $dr_*=dr/\sqrt{AB}$.
The effective potential reads
\begin{eqnarray}\label{f4}
    V_{s,l}=A\frac{l(l+1)}{r^2}+\frac{1}{2r}\left(A'B+AB'\right).
\end{eqnarray}

\subsection{Boundary Conditions}

There are freedoms in choosing boundary conditions depending on
physical pictures of different problems. We will consider the
plane wave scattering in our work.

\subsubsection{Boundary Condition at Infinity}
In the remote region, by virtue of the scattering problem, the
scalar field at infinity is in the form
\begin{equation}\label{f5}
    \Phi \sim \left[e^{i\omega r\cos\theta}+\frac{f(\theta)}{r}e^{i\omega
    r}\right]e^{-i\omega t}.
\end{equation}
Referring to the scheme of partial wave method in quantum
mechanics, the plane wave can be expanded into spherical waves as,
\begin{equation}\label{f6}
    e^{i\omega r\cos\theta}=\sum_{l=0}^{\infty}(2l+1)i^lj_l(\omega
    r)P_l(\cos \theta).
\end{equation}
When $r$ tends to infinity,
\begin{equation}\label{f7}
    e^{i\omega
    r\cos\theta}\sim\sum_{l=0}^{\infty}(-1)^{l+1}\frac{2l+1}{2\omega r}[e^{-i\omega r}-(-1)^le^{i\omega
    r}]P_l(\cos\theta).
\end{equation}
The scattered wave can also be expanded in the form
\begin{equation}\label{f8}
    \frac{f(\theta)}{r}e^{i\omega
    r}=\sum_{l=0}^{\infty}(2l+1)f^s_l\frac{e^{i\omega r}}{2\omega r}
    P_l(\cos\theta),
\end{equation}
where $f^s_l$ are complex constants, and the superscript "s"
indicates the case of scalar field. Inserting this equation
together with Eq. (\ref{f7}) into Eq. (\ref{f5}), we have the
boundary condition at infinity
\begin{equation}\label{f9}
    \Phi\sim e^{-i\omega t}\sum_{l=0}^{\infty}(-1)^{l+1}\frac{2l+1}{2\omega r}
    [e^{-i\omega r_*}-(-1)^lS^s_le^{i\omega
    r_*}]P_l(\cos\theta),
\end{equation}
where $S^s_l=1+f^s_l$, and the radius $r$ on the exponentials are
replaced by tortoise coordinate $r_*$ for future convenience. Note
that at infinity $dr_*\sim dr$ so that such replacement on the
exponentials makes no difference. In accordance, we have the
boundary condition of $R_l$ at infinity
\begin{equation}\label{f10}
    R_l\sim(-1)^{l+1}\frac{2l+1}{2\omega r}[e^{-i\omega r_*}-(-1)^lS^s_le^{i\omega
    r_*}].
\end{equation}

\subsubsection{Boundary Condition at Horizon}
Now we examine the boundary condition in the near-horizon region.
We can express the solution of  Eq. (\ref{f3}) in power series \cite{persides}
\begin{equation}\label{f11}
    R_l=(r/r_H-1)^{\rho}\sum_{n=0}^{\infty}d_{l,n}(r/r_H-1)^n.
\end{equation}
In the vicinity of the horizon, we have the asymptotic behavior
\begin{equation}\label{f12}
    \sqrt{AB}\sim\sqrt{A_1B_1}(r/r_H-1),
\end{equation}
as well as
\begin{equation}\label{f13}
    V_{s,l}(r)\sim O(r-r_H),  \ \ \ (r\rightarrow r_H+0).
\end{equation}
Inserting the expansions Eq. (\ref{f12},\ref{f13})  into the radial
equation Eq. (\ref{f3}), we have
\begin{equation}\label{index}
    A_1B_1[\rho(\rho-1)+\rho]+r_H^2\omega^2=0,
\end{equation}
or,
\begin{equation}\label{f14}
    \rho=\pm\frac{i\omega r_H}{\sqrt{A_1B_1}}.
\end{equation}
Since nothing can escape from the black hole, only the minus sign is
acceptable. Thus we have the boundary condition of $R_l(r)$ near
the horizon\cite{persides}
\begin{equation}\label{f15}
    R_l(r)\sim g_l(r/r_H-1)^{-i\omega r_H/\sqrt{A_1B_1}},
\end{equation}
where $g_l$ are the zero-th coefficient of the expansion of $R_l$:
$g_l=d_{l,0}$, which is dimensionless. The boundary condition for
the total perturbational scalar field $\Phi$ is given by
\begin{equation}\label{f16}
    \Phi\sim e^{-i\omega t}\sum_{l=0}^{\infty} g_l(r/r_H-1)^{-i\omega
    r_H/\sqrt{A_1B_1}}P_l(\cos\theta).
\end{equation}

\subsection{The Energy Flux and the Absorption Cross Section}

The energy flux is derived from the energy-momentum tensor of the
scalar field. For massless scalar field, it is generally expressed
as \cite{early9}
\begin{equation}\label{f17}
    T_{\mu\nu}=\Phi_{;\mu}\Phi^*_{;\nu}
    -\frac{1}{2}g_{\mu\nu}\Phi^{;\alpha}\Phi^*_{;\alpha}.
\end{equation}
The energy-momentum flow of the spacetime is defined by\cite{early4}
\begin{equation}\label{f18}
    P^{\mu}=T^{\mu}_{\nu}\xi^{\nu}(t),
\end{equation}
where vector $\xi(t)=\partial/\partial t$ is the time-like Killing
vector. We now calculate the energy falling into the black hole
per unit time using Gauss' theorem. In the theorem, for any
appropriate four vector $A^{\mu}$ and space time sector
$\mathcal{N}$, there holds
\begin{equation}\label{f19}
    \int_\mathcal{N}A^{\mu}_{;\mu}\mathbf{\Omega}_4
    =\int_{\partial\mathcal{N}}A^{\mu}n_{\mu}\mathbf{\Omega}_3,
\end{equation}
where $\mathbf{\Omega_4}$ and $\mathbf{\Omega_3}$ are the volume
elements of the spacetime sector $\mathcal{N}$, and its boundary
$\partial\mathcal{N}$.  $n^{\mu}$ is the normal vector of the
boundary $\partial \mathcal{N}$ which satisfies
$\hat{\mathbf{n}}\wedge\mathbf{\Omega}_3=\mathbf{\Omega}_4$, where
$\hat{\mathbf{n}}$ is the 1-form physically equivalent to vector
$n^{\mu}$. Here we may choose the space time sector $\mathcal{N}$
to be the spacial area within two spheres of radius $r_1$ and
$r_2$, and with the time interval between $t_1$ and $t_2$, whose
boundaries are $\partial\mathcal{N}$. To find out the energy
falling into the hole, we focus on the hypersurface of $r=r_1$,
whose normal vector is $n_{\mu}=(0,\sqrt{A/B},0,0)$, and we let
$r_1\rightarrow r_H$. Thus the energy falling into the hole within
the time interval between $t_1$ and $t_2$ is \cite{Ch},
\begin{equation}\label{f20}
    E^{(abs)}=\int^{t_2}_{t_1}dt\int_{4\pi}P^{\mu}n_{\mu}
    r_H^2\sin\theta d\theta
    d\varphi=\int^{t_2}_{t_1}dt\int_{4\pi}T_{0\mu}n^{\mu}
    r_H^2\sin\theta d\theta d\varphi.
\end{equation}\label{f21}
Applying the explicit forms of $T_{0\mu}$ and $n^{\mu}$, as well
as the boundary condition Eq. (\ref{f16}), we have, at the horizon
\begin{eqnarray}\label{f22}
    T_{0\mu}n^{\mu}=-\sqrt{AB}\
    \partial_t\Psi\partial_r\Psi^*|_{r_H}
    =\omega^2\left|\sum_{l=0}^{\infty}g_lP_l(\cos\theta)\right|^2.
\end{eqnarray}
Inserting this into Eq. (\ref{f20}), we have the total energy
falling into the horizon per unit time:
\begin{equation}\label{f23}
    \Phi^{(abs)}=\frac{dE^{(abs)}}{dt}=4\pi\omega^2r_H^2
    \sum_{l=0}^{\infty}\frac{|g_l|^2}{2l+1}.
\end{equation}
Here, we have used the orthogonality of Legendre polynomials,
\begin{equation}\label{f26}
    \int_{-1}^{1}{P_l(x)P_{l'}(x)dx}=\frac{\delta_{ll'}}{2l+1}.
\end{equation}

On the other hand, following the similar procedure as above, we
can derive the expression of the energy flux of the incoming wave
 at infinity, which is simply
\begin{equation}\label{f24}
    j^{(inc)}=\omega^2.
\end{equation}
Therefore for the massless scalar wave the total energy absorption
by the black hole is given by
\begin{eqnarray}\label{f25}
    \sigma^{abs}_{s}=\frac{\Phi^{(abs)}}{j^{(inc)}}=4\pi r_H^2
    \sum_{l=0}^{\infty}\frac{|g_l|^2}{2l+1}.
\end{eqnarray}
For angular index $l$, the partial absorption cross section is
expressed as
\begin{equation}\label{f27}
    \sigma_{s,l}^{abs}=4\pi r_H^2 \frac{|g_l|^2}{2l+1}.
\end{equation}
After we get $g_l$ we can obtain the final result of the
absorption spectrum.

For the radial equation Eq. (\ref{f3}), the Wronskians for any two
solutions $R_l^{(1)}$ and $R_l^{(2)}$ is
\begin{equation}\label{SWs}
    W[R_l^{(1)},R_l^{(2)}]=\frac{K_{12}}{r^2\sqrt{AB}},
\end{equation}
with the constant $K_{12}$ to be determined by the explicit form
of the two solutions. Using the asymptotic solution of $R_l$ at
infinity Eq. (\ref{f10}), we have
\begin{equation}\label{SWsInf}
    W[R_l^*, R_l]=-\frac{i(2l+1)^2}{2\omega r^2
    \sqrt{AB}}\left(1-|S^s_l|^2\right),
\end{equation}
and employing the asymptotic solution at the horizon Eq.
(\ref{f15}), we get
\begin{equation}\label{SWsHor}
    W[R_l^*, R_l]=-\frac{2i\omega r_H^2}{r^2
    \sqrt{AB}}|g_l|^2.
\end{equation}
Equating the above two equations, we find
\begin{equation}\label{f28}
    |g_l|^2=\frac{(2l+1)^2}{4\omega^2 r_H^2}\left(1-|S^s_l|^2\right).
\end{equation}
Thus with Eq. (\ref{f27}) we can express the energy absorption in
the form
\begin{equation}\label{SOT}
    \sigma^{abs}_{s,l}=\frac{\pi}{\omega^2}(2l+1)\Gamma^s_l,
\end{equation}
where $\Gamma^s_l$ is defined in terms of $S^s_l$ as
\begin{equation}\label{f29}
    \Gamma^s_l=1-|S_l^s|^2.
\end{equation}

\begin{figure}
  \includegraphics[width=9cm]{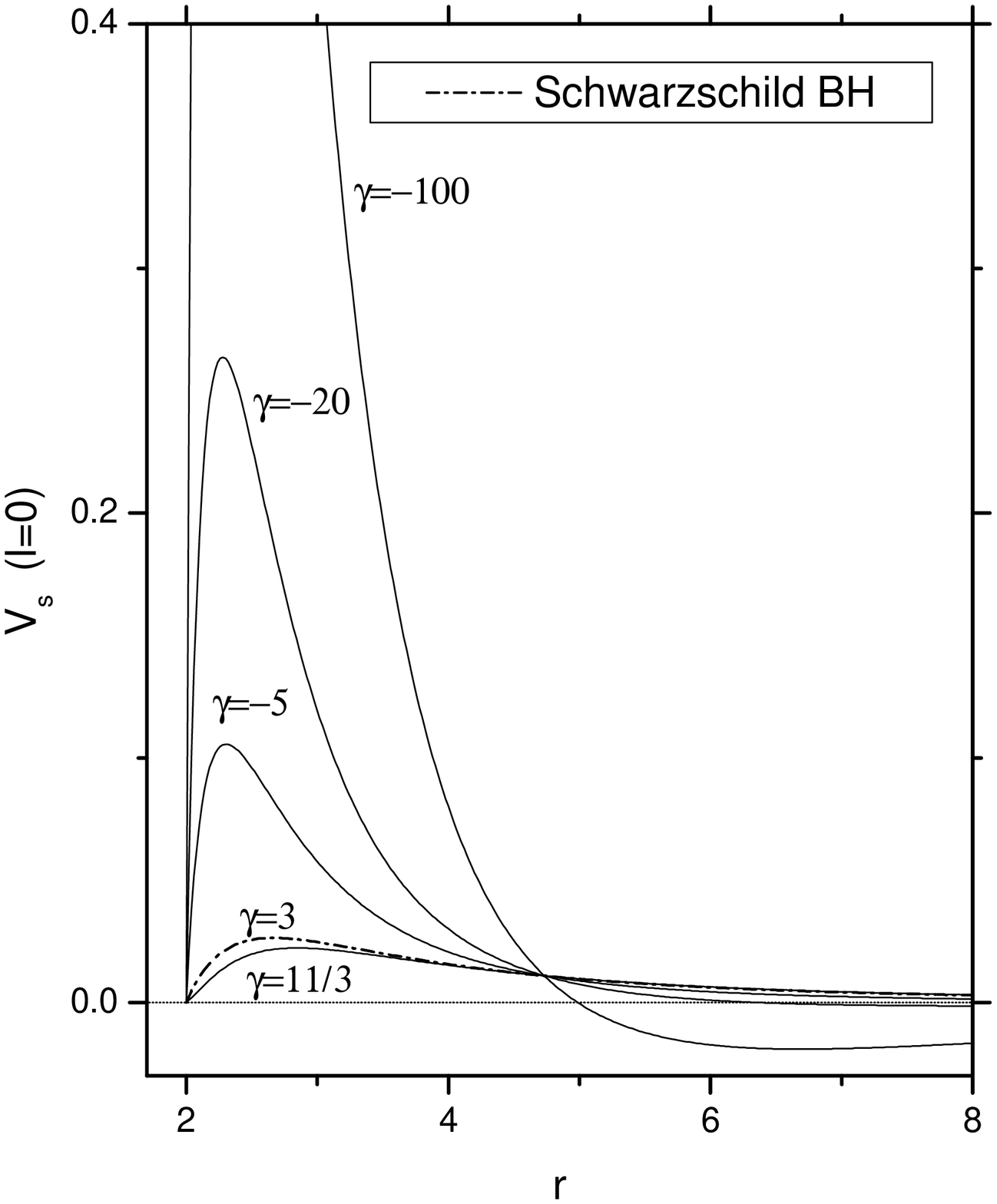}\includegraphics[width=9cm]{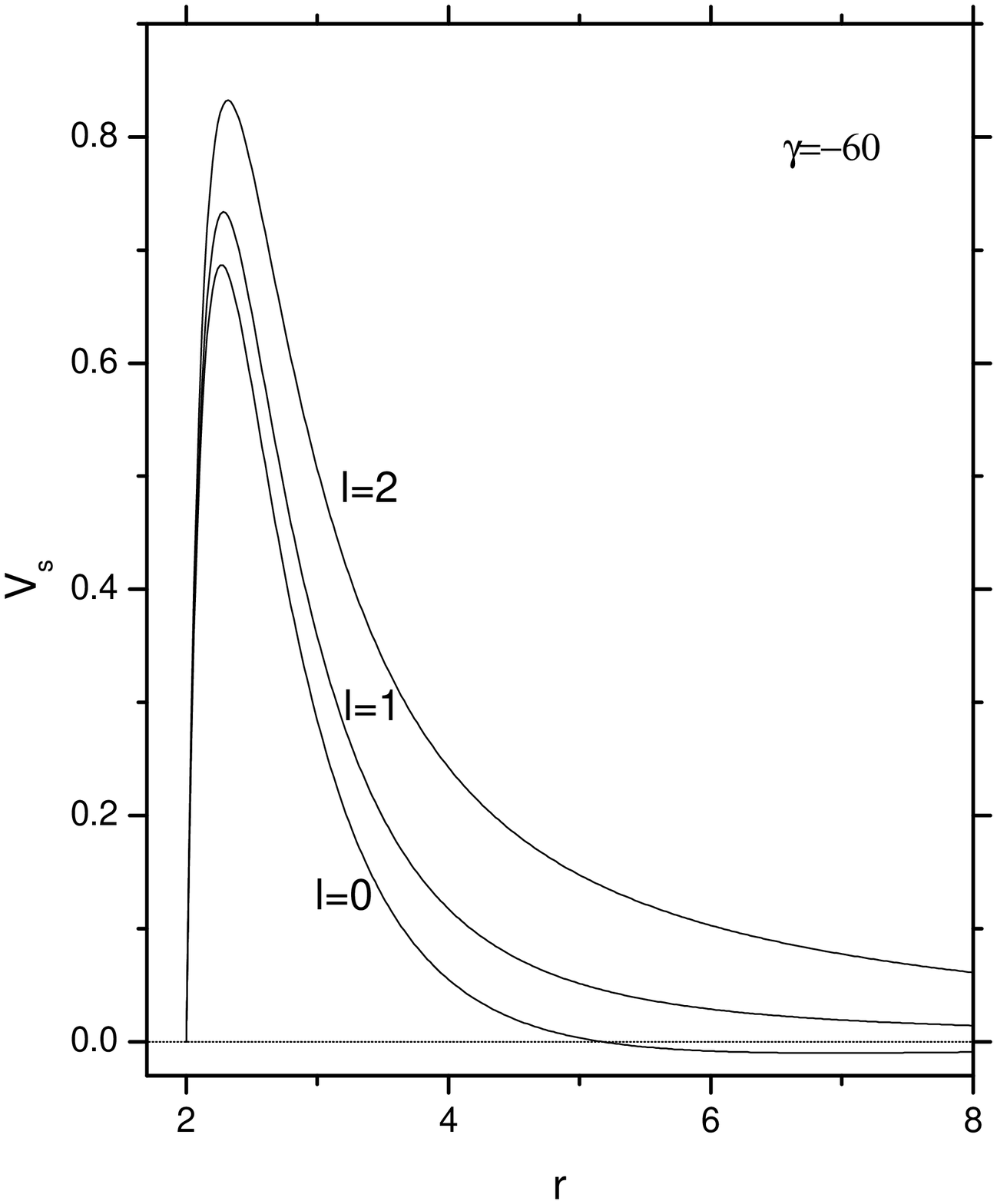}\\
  \caption{The effective potential of the scalar perturbation with the change of $\gamma$ and $l$.}\label{VsGm}
\end{figure}

\begin{figure}
  \includegraphics[width=9cm]{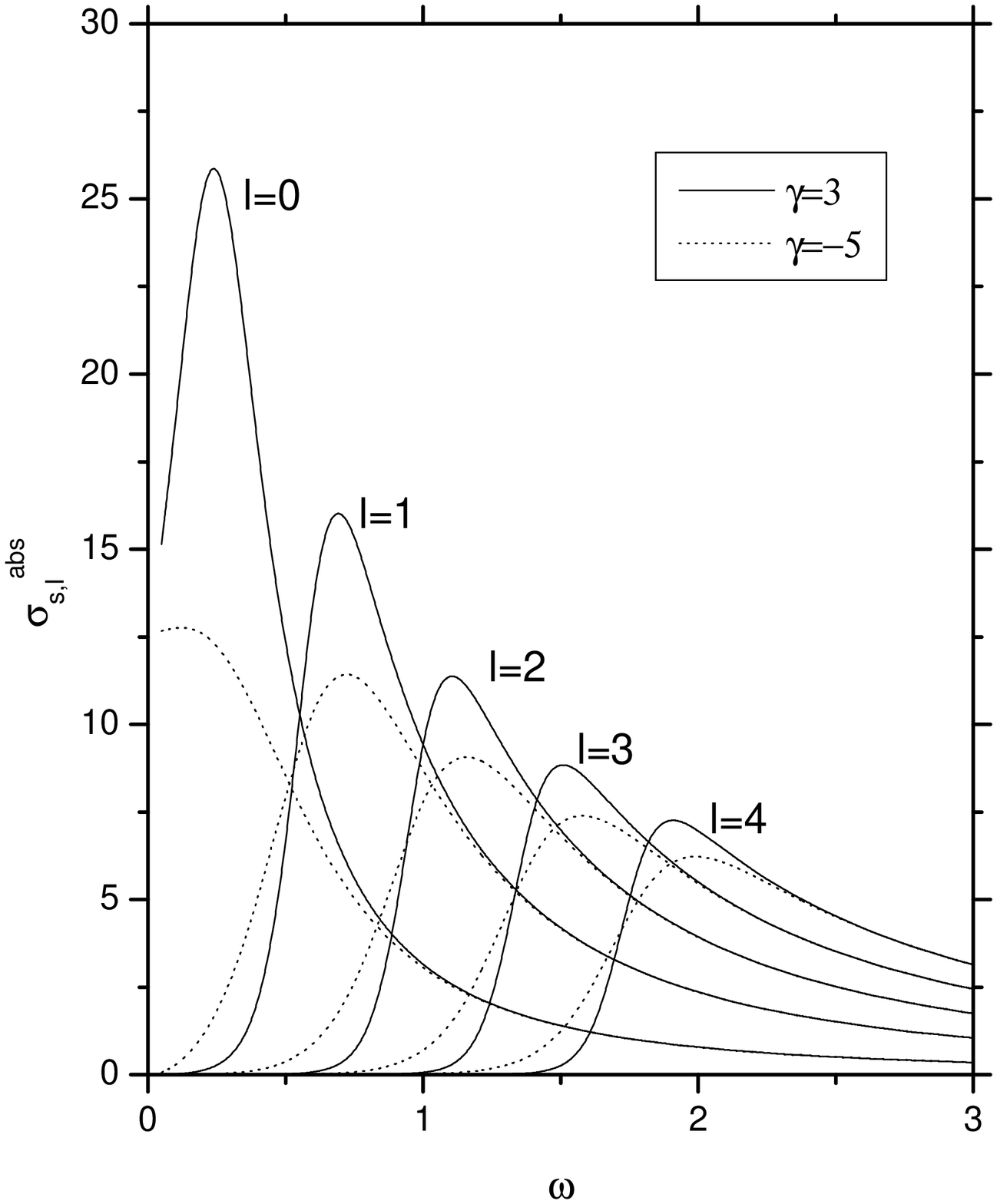}\includegraphics[width=9cm]{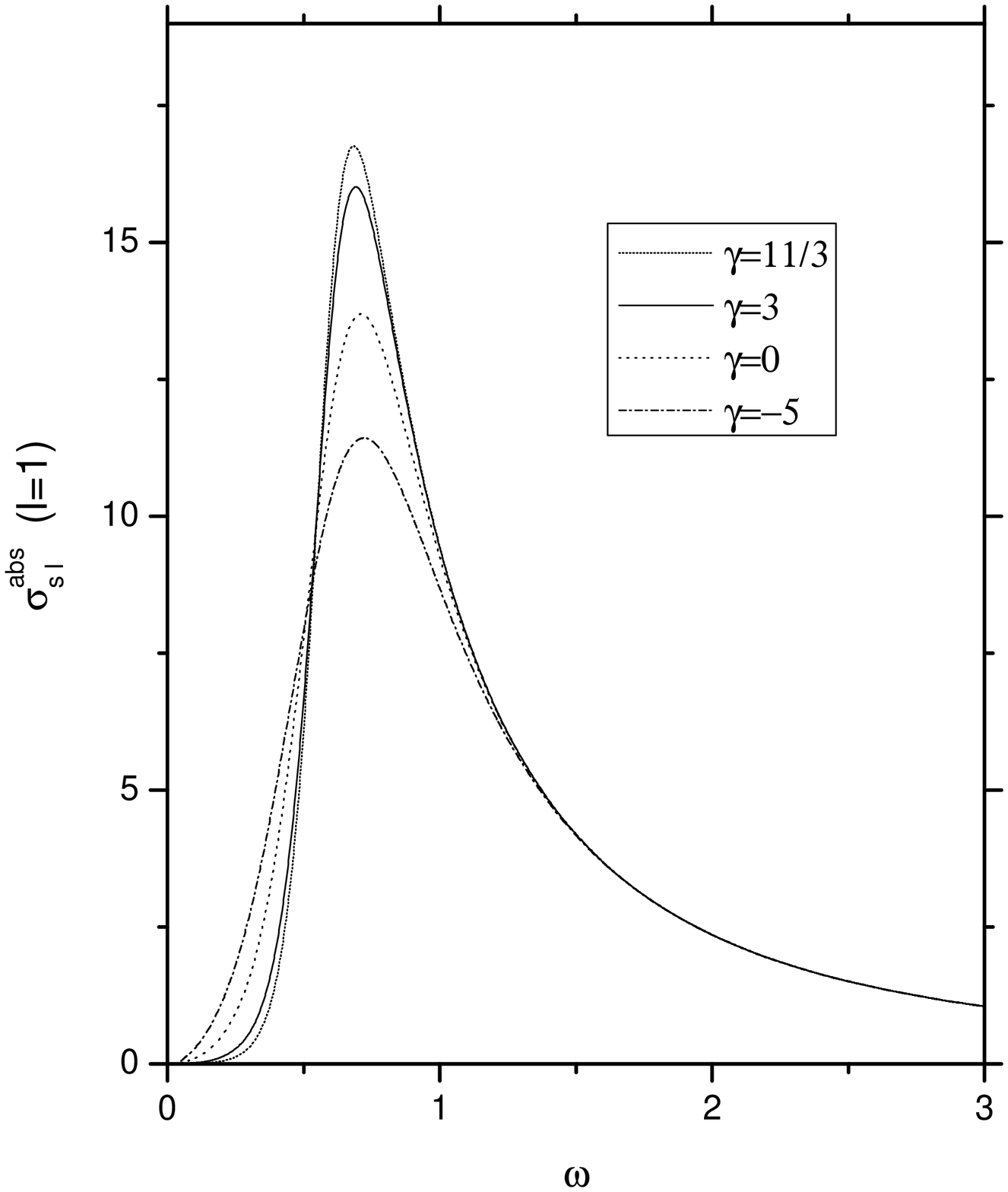}\\
  \caption{The partial energy absorption of the scalar wave with the change of $\gamma, l$}\label{Vsl0}
\end{figure}

\begin{figure}
  \includegraphics[width=12cm]{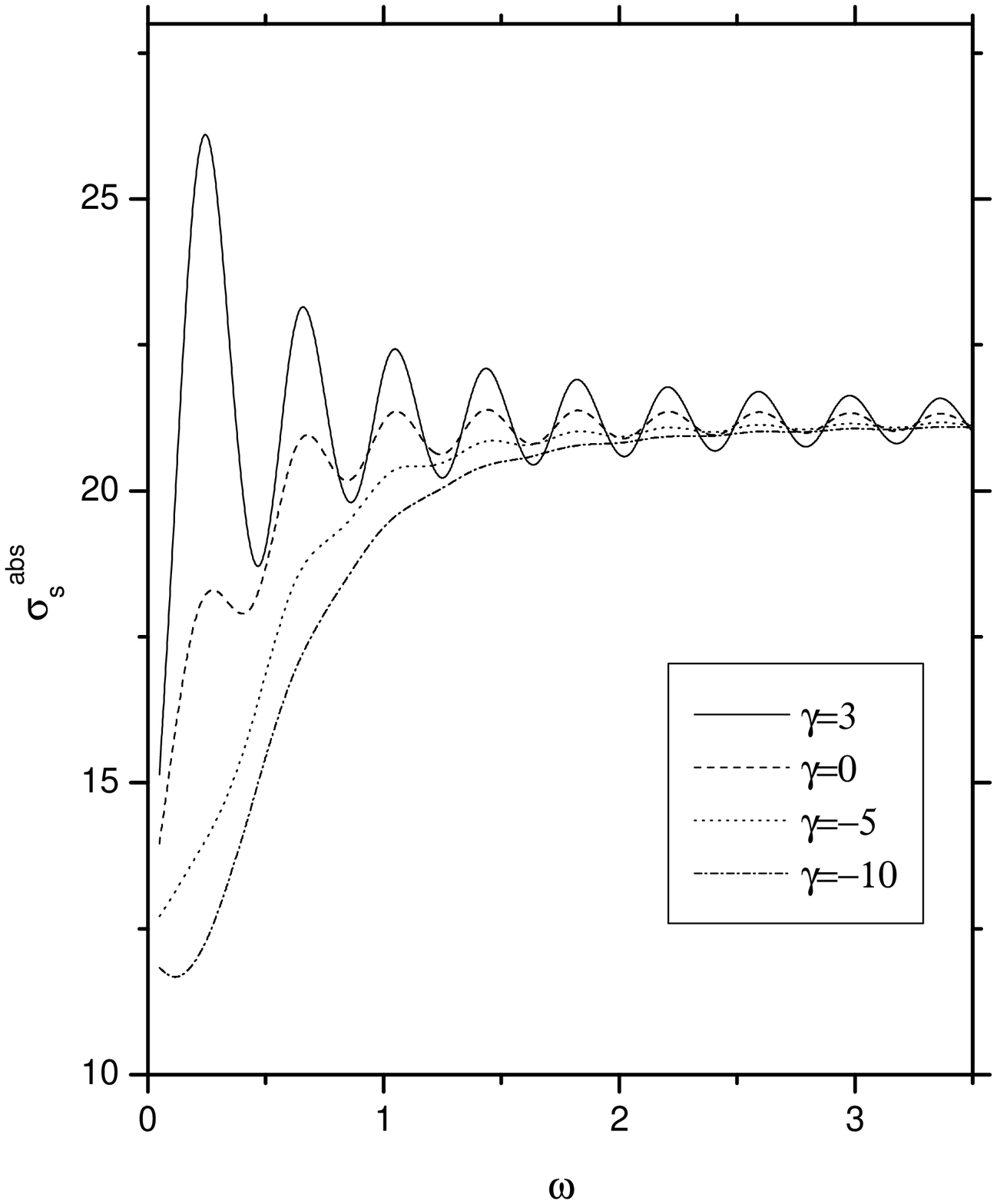}\\
  \caption{The total energy absorption of the scalar wave.}
   \label{totS}
\end{figure}

\subsection{Numerical Study for the Scalar Perturbation}

Now we proceed to calculate numerically the coefficients $g_l$.
Instead of solving Eq. (\ref{f3}) directly for
$R_l$, we choose the particular solution $\varphi_l(r)$,
normalized as
\begin{equation}\label{snm1001}
    \varphi_l(r)=g_l^{-1}R_l(r).
\end{equation}
In the asymptotically flat region, it resolves
into outgoing and ingoing waves
\begin{equation}\label{snm3}
    \varphi_l(r)=f_l^{(-)}\mathcal{F}_{l(+)}+f_l^{(+)}\mathcal{F}_{l(-)},
\end{equation}
where\cite{park,persides}
\begin{equation}\label{snm2}
    \mathcal{F}_{l(\pm)}=e^{\mp i\omega
    r_*}\sum_{n=0}^{\infty}\tau^{(\pm)}_{l,n}\left(\frac{r_H}{r}\right)^{n+1},
\end{equation}
satisfying $\mathcal{F}_{l(+)}=\mathcal{F}_{l(-)}^*$ and the
coefficients $\tau_{l,0}^{(\pm)}$ being unity for simplicity. The
combination coefficients $f_l^{(\pm)}$ are called jost functions.
Comparing this to Eq. (\ref{f10}), we have
\begin{eqnarray}\label{snm4}
    g_l=\frac{(-1)^{l+1}}{\omega}\frac{l+1/2}{f_l^{(-)}},\\
    \label{snm50}
    S_l^s=(-1)^{l+1}\frac{f_l^{(+)}}{f_l^{(-)}}.\ \ \
\end{eqnarray}
Thus if the jost functions are known, we can get the solution to
our problem. Applying Eq. (\ref{SWs}) to the two particular
solutions at infinity, we have
\begin{eqnarray}\label{snm5}
    W[\mathcal{F}_{l(+)},\mathcal{F}_{l(-)}]
    =\frac{2i\omega r_H^2}{r^2\sqrt{AB}},
\end{eqnarray}
and therefore we have the jost functions
\begin{equation}\label{snm6}
    f_l^{(\pm)}=\mp W[\varphi_l, \mathcal{F}_{l(\pm)}]\frac{r^2\sqrt{AB}}{2i\omega
    r_H^2}.
\end{equation}
In numerical calculation, we would like to express
$g_l$ in Eq. (\ref{f25}) and Eq. (\ref{f27}) in terms of the jost
functions, and the absorption cross section therefore reads
\begin{equation}\label{snm7}
    \sigma^{abs}_{s,l}=\frac{\pi}{\omega^2}
    \frac{2l+1}{|f_l^{(-)}|^2},
\end{equation}
and
\begin{equation}\label{snm8}
    \sigma^{abs}_s=\frac{\pi}{\omega^2}\sum_{l=0}^{\infty}
    \frac{2l+1}{|f_l^{(-)}|^2}.
\end{equation}

We now report our numerical result for the massless scalar
perturbation in the CFM braneworld black hole background.
Inserting the metric components into Eq. (\ref{f4}), the effective
potential has the form
\begin{equation}\label{pots}
    V_{s,l}(r)=\left(1-\frac{2M}{r}\right)\left[\frac{l(l+1)}{r^2}
    +\frac{2M}{r^3}+\frac{M(\gamma-3)(r^2-6Mr+6M^2)}{r^3(2r-3M)^2}\right].
\end{equation}
Taking $M=1$, the bahavior of the potential is shown in Fig.1. We
see that there is a potential barrier outside the horizon and for
the fixed angular index $l$ this barrier increases for more
negative $\gamma$. Fixing $\gamma$, we find that the potential
barrier increases with the increase of the angular index. When
$\gamma$ becomes small, the potential has a negative well after
the positive barrier. The negative well appears earlier for very
negative $\gamma$. This was also observed in \cite{eabdalla}.
Comparing with the height of the positive barrier, the absolute
value of the negative peak is very small for chosen $\gamma$. Thus
the potential barrier will dominate in the determination of the
wave dynamics, which qualitatively leads to the general
quasinormal ringing as disclosed in \cite{eabdalla}.

The spectrum of absorption cross sections for both Schwarzschild
and the brane-world case with $\gamma=-5$ are shown in  Fig.2a for
different $l$. Fig.2b displays the partial absorption cross
section for chosen $l$ and different $\gamma$. It is noticed that
the absorption cross section of the brane-world black hole can be
either smaller or larger than that in the Schwarzschild black hole
provided that $\gamma$ is smaller or bigger than $3$. This result
is consistent with the behavior of the potential barriers as shown
in Fig.1. The higher or lower barrier of the potential could
decrease or increase the absorption of the scalar field around the
brane-world black hole. For the low $\omega$ region, we observed
that the absorption is higher for smaller values of $\gamma$. This
could be the imprint of the negative well in the potential, since
different from the barrier, the potential well enhances the
absorption rate. But the overall behavior of the energy absorption
shows that the potential barrier dominates the physics.

In Fig.3, we show the spectrum of the total absorption cross
section, we see that with the decrease of $\gamma$, there are
fewer oscillations. For smaller $\gamma$, in the low energy band
the total absorptions are suppressed. This is because of the
increase in the potential barrier which plays the dominate role
for smaller $\gamma$ leading to the decrease of the absorption of
the incoming wave. The full penetration occurs when the incoming
wave with very high energy, which explains why the total
absorption approaches a constant value at high frequency.

\section{Gravitational Perturbation and Its Energy Absorption Cross Section}
Now we start the formulation of gravitational perturbation,
of which the detailed deduction requires special techniques
provided in \cite{Ch, a1}. We will concentrate our attention here
on the brane-localized axial perturbation in the background of the
braneworld black hole.

\subsection{Axial Gravitational Perturbation Equations}

Using the formalism introduced in \cite{Ch},  the axially
perturbed metric Eq. (\ref{e3}) is
\begin{equation}\label{g1}
    ds^2=e^{2\nu}(dt)^2-e^{2\psi}(d\varphi-\varpi dt-q_2 dx^2-q_3
    dx^3)^2-e^{2\mu_2}(dx^2)^2-e^{2\mu_3}(dx^3)^2.
\end{equation}
Here
\begin{equation}\label{g2}
    x^2=r,\ \ x^3=\theta,
\end{equation}
$\varpi$ and $q_i\ (i=2,3)$ are quantities that embody the
perturbation. Due to the lack of knowledge of the 5D bulk metric,
we restrict to the brane localized perturbation governed by the
perturbation equation
\begin{equation}\label{g3}
    \delta R_{\mu\nu}=0.
\end{equation}
This simplification is supported by the analysis of gravitational
shortcuts \cite{abdalla2} which shows that gravitational fields do not
travel deep into the bulk. It can be justified at least in a
regime where the perturbation energy does not exceed the threshold
of the Kaluza-Klein massive modes. By defining
\begin{equation}\label{g4}
    Q_{ij}=q_i,_j-q_j,_i,\ \  Q_{0i}=\varpi,_i-q_i,_0,
\end{equation}
one finds equations
\begin{equation}\label{g5}
    (e^{3\psi+\nu-\mu_2-\mu_3}Q_{23}),_{3}=-e^{3\psi-\nu+\mu_3-\mu_2}Q_{02},_{0}
\end{equation}
from $\delta R_{12}=0$, and
\begin{equation}\label{g6}
    (e^{3\psi+\nu-\mu_2-\mu_3}Q_{23}),_2=e^{3\psi-\nu+\mu_2-\mu_3}Q_{03},_0
\end{equation}
from $\delta R_{13}=0$. We take
\begin{equation}\label{g7}
    e^{3\psi+\nu-\mu_2-\mu_3}Q_{23}=r^2\sqrt{AB}\sin^3\theta Q_{23}=Q
\end{equation}
in Eq. (\ref{g5}) and Eq. (\ref{g6}), and then cancel $\varpi$, to
arrive at an equation of $Q$:
\begin{equation}\label{g8}
    -r^2\frac{\partial^2 Q}{\partial t^2}+r^4\sqrt{AB}\frac{\partial}
    {\partial r}\left(\frac{\sqrt{AB}}{r^2}\frac{\partial Q}{\partial
    r}\right)+A\sin^3 \theta \frac{\partial}{\partial
    \theta}\left(\frac{1}{\sin^3 \theta}\frac{\partial Q}{\partial
    \theta}\right)=0.
\end{equation}
Analogously, this equation plays the same role as the Klein-Gorden
equation in the scalar perturbation. However the equation here
cannot be separated in terms of Legendre polynomials as in the
Klein-Gorden equation, the Gegenbauer polynomials are employed to
carry out the separation\cite{Ch}. $Q$ is now seperated as
\begin{equation}\label{facQ}
    Q(t,r,\theta)=\sum_{l=2}^{\infty}rZ_l(r)C^{-3/2}_{l+2}(\cos\theta)e^{-i \omega t},
\end{equation}
where the Gegenbauer polynomials $C^{-3/2}_{l+2}(\theta)$ satisfy
the equation
\begin{equation}\label{g9}
    \sin^{2\nu}\theta
    \frac{d}{d\theta}\left(\sin^{-2\nu}\theta
    \frac{dC^{\nu}_l(\cos\theta)}{d\theta}\right)+l(l+2\nu)C^{\nu}_l(\cos\theta)=0.
\end{equation}
The radial Schr\"{o}dinger-like equation can be derived in the
form,
\begin{equation}\label{g10}
    \frac{d^2Z_l}{dr_*^2}+[\omega^2-V_{g,l}(r)]Z_l=0.
\end{equation}
$r_*$ is the tortoise coordinate defined by $dr_*=dr/\sqrt{AB}$,
and the effective potential is
\begin{eqnarray}\label{g11}
    V_{g,l}(r)=A\frac{(l-1)(l+2)}{r^2}+\frac{2AB}{r^2}-\frac{A'B+B'A}{2r}.
\end{eqnarray}

\subsection{Boundary Conditions}

We set the boundary conditions to be those of plane wave scattering.

\subsubsection{Boundary Condition at Infinity}

It is expected that the quantity $Q=r^2\sqrt{AB}\sin^3\theta
Q_{23}$ has the similar behavior of the scattered spherical wave
since a straightforward examination shows that
$Q(t,r,\theta)=e^{i(\omega r \cos\theta-\omega t)}$ is exactly a
solution of Eq. (\ref{g8}) in the region far from the black hole.
Therefore it is appropriate to write
\begin{equation}\label{g12}
    Q(t,r,\theta)\sim \left[e^{i\omega r \cos \theta}
    +\frac{f(\theta)}{r}e^{i\omega r}\right]e^{-i\omega t}, \ \
    r\rightarrow \infty.
\end{equation}
For the infinite $r$, $Q$ behaves as Eq. (\ref{g12}), which can
also be expanded in a combination of Gagenbauer polynomials as the
decomposition of $Q$ in Eq. (\ref{facQ}).  The expansion of the
plane wave part is \cite{ToI}
\begin{equation}\label{g13}
    e^{i\omega r\cos \theta}=-\frac{1}{3}\sum_{l=-2}^{\infty}(l+1/2)i^l(\omega r)^2
    j_l(\omega r)C^{-3/2}_{l+2}(\cos\theta).
\end{equation}
Note that in practice we calculate only the partial absorption
cross section of low angular indices, thus when $\omega r$ is a
large number, the asymptotic approximation for $j_l(\omega r)$
reads
\begin{equation}\label{g14}
    j_l(x)\sim \frac{1}{x}\sin(x-l\pi/2)=-\frac{i^l}{2x}\left[e^{-ix}-(-1)^le^{ix}\right]
\end{equation}
for $x\gg 1$ and $x\gg l$, where $x=\omega r$. There exists a
certain $l_0$ and before this $l_0$ the summation terms of Eq.
(\ref{g13}) may be replaced by the asymptotic form,
\begin{eqnarray}\label{g15}
    e^{i\omega r\cos
    \theta}=\frac{1}{6}\sum_{l=-2}^{l_0}(l+1/2)(-1)^l(\omega
    r)\left[e^{-i\omega r}-(-1)^le^{i\omega
    r}\right]C^{-3/2}_{l+2}(\cos\theta) \nonumber \\
    +\left[-\frac{1}{3}\sum_{l=l_0+1}^{\infty}(l+1/2)i^l(\omega r)^2
    j_l(\omega r)C^{-3/2}_{l+2}(\cos\theta)\right].\ \ \ \ \ \
\end{eqnarray}
The larger $\omega r$ is, the larger $l_0$ will be.

We also need  the expansion of the term representing the
scattered wave. Formally we may write
\begin{equation}\label{g16}
    \frac{f(\theta)}{r}e^{ikr}=\sum_{l=-2}^{\infty} \chi_l(r)
    C^{-3/2}_{l+2}(\cos \theta).
\end{equation}
Similar to the plane wave, we expect that before some integer
$l_0'$, the summation coefficients in Eq. (\ref{g16}) are
asymptotically
\begin{equation}\label{g17}
    \chi_l(r)\sim \frac{1}{6}(l+1/2)(\omega r)f^g_le^{i\omega
    r},
\end{equation}
at some very large $\omega r$, where $f^g_l$ are complex numbers
depending solely on index $l$. Thus we can write the expansion Eq.
(\ref{g16}) as
\begin{eqnarray}\label{g18}
    \frac{f(\theta)}{r}e^{i\omega r}=\frac{1}{6}\sum_{l=-2}^{l_0'}
    (l+1/2)(\omega r)f^g_le^{i\omega
    r}C_{l+2}^{-3/2}(\cos \theta)\nonumber \\
    +\sum_{l=l_0'+1}^{\infty}\chi_l(r)
    C^{-3/2}_{l+2}(\cos \theta).\ \ \ \ \ \ \
\end{eqnarray}
Combining Eq. (\ref{g15}) and Eq. (\ref{g18}), we have the
applicable form of the asymptotic behavior of $Q(t,r,\theta)$:
\begin{equation}\label{g19}
    Q\sim \frac{e^{-i\omega t}}{6}\sum_{l=2}^{\min(l_0,l_0')}(l+1/2)(-1)^l(\omega
r)\left[e^{-i\omega r}-(-1)^lS^g_le^{i\omega
r}\right]C^{-3/2}_{l+2}(\cos\theta)+\ldots.
\end{equation}
Here the omitted terms are those over the upper limit for applying
the asymptotic form, which are not applicable to our numerical
calculation. $S^g_l=1-f^g_l$ is a complex number being the ratio
of incoming and outgoing wave. The summation starts from index 2
for the modes from $l=-2$ to $l=1$ are not physical in
gravitational perturbation. Since $\min(l_0,l_0')$ certainly
exceeds our need for numerical calculation, we may write for
simplicity
\begin{equation}\label{g20}
    Q\sim \frac{e^{-i\omega t}}{6}\sum_{l=2}^{\infty}(l+1/2)(-1)^l(\omega
r)\left[e^{-i\omega r_*}-(-1)^lS^g_le^{i\omega
r_*}\right]C^{-3/2}_{l+2}(\cos\theta),
\end{equation}
which is valid for a limited number of angular indices. As done in
the scalar case, in the exponentials we have replaced $r$ into
$r_*$, since in the remote region the spacetime is flat and we
actually have $dr_*\sim dr$. Therefore the boundary condition for
$Z_l(r)$ at infinity reads
\begin{equation}\label{g21}
    Z_l(r)\sim \frac{1}{6}(l+1/2)(-1)^l\omega
    \left[e^{-i\omega r_*}-(-1)^lS^g_le^{i\omega r_*}\right]\ \
    r\rightarrow \infty.
\end{equation}



\subsubsection{Boundary Condition at the Horizon}
We now study the boundary condition in the vicinity of the event
horizon $r_H$. Expanding $Z_l(r)$ near the horizon\cite{persides}
\begin{equation}\label{g99}
    Z_l(r)=\left(\frac{r}{r_H}-1\right)^{\rho}
    \sum_{n=0}^{\infty}c_{l,n}\left(\frac{r}{r_H}-1\right)^n,
\end{equation}
we have
\begin{equation}\label{g22}
    \frac{dZ_l}{dr}\sim \frac{p_l \rho}{r_H}
    \left(\frac{r}{r_H}-1\right)^{\rho-1}, \ \ \ \
    \frac{d^2Z_l}{dr^2}\sim \frac{p_l
    \rho(\rho-1)}{r_H^2}\left(\frac{r}{r_H}-1\right)^{\rho-2},
\end{equation}
when $r\rightarrow r_H$.  $p_l=c_{l,0}$ with the
dimension of length$^{-1}$. Considering Eq.(\ref{f12}),  in the
vicinity of the event horizon Eq. (\ref{g22}) becomes
\begin{eqnarray}\label{g23}
    \frac{d^2Z_l}{dr_*^2}\sim
    A_1B_1\left(\frac{r}{r_H}-1\right)^2\frac{d^2Z_l}{dr^2}
    +\frac{A_1B_1}{r_H}\left(\frac{r}{r_H}-1\right)\frac{dZ_l}{dr}
    \sim \frac{p_l \rho^2 (A_1B_1)}{r_H}\left( \frac{r}{r_H}-1
    \right)^{\rho},
\end{eqnarray}
and
\begin{equation}\label{g24}
    [\omega^2-V_{eff}(r)]Z_l\sim p_l \omega^2
    \left(\frac{r}{r_H}-1\right)^{\rho}.
\end{equation}
Inserting Eq. (\ref{g23}) and Eq. (\ref{g24}) into Eq.
(\ref{g10}), we obtain
\begin{equation}\label{indexgrav}
    A_1B_1[\rho(\rho-1)+\rho]+r_H^2\omega^2=0,
\end{equation}
where
\begin{equation}\label{g25}
    \rho=\pm i\omega r_H/\sqrt{A_1B_1}.
\end{equation}
Similar to the scalar case,  we choose the minus sign in the
$\rho$ expression since there is only ingoing wave near the event
horizon and nothing can escape from the black hole. The
boundary condition at the horizon then reads
\begin{equation}\label{g26}
    Z_l(r)\sim p_l(r/r_H-1)^{-i\omega r_H/\sqrt{A_1B_1}},\ \
    r\rightarrow r_H+0.
\end{equation}
In accordance the behavior of $Q(t,r,\theta)$ at the horizon is
\begin{equation}\label{g27}
    Q(t,r,\theta)\sim e^{-i\omega t}r_H(r/r_H-1)^{-i\omega r_H/\sqrt{A_1B_1}}
    \sum_{l=2}^{\infty}p_lC^{-3/2}_{l+2}(\cos\theta)\ ,\ \
    r\rightarrow r_H+0.
\end{equation}

\subsection{Application of the Newman-Penrose Formalism}

The study of the gravitational perturbation and the gravitational
energy absorption spectrum involves considerable algebraic
complexity. There is a well developed approach to this problem
provided by Teukolsky \cite{early4, early9}, where Newman-Penrose
formalism was employed. In this section we will employ the
formalism to calculate the energy flux coming from infinity and
the energy falling into the black hole.

\subsubsection{Formalism in the Kinnersley Tetrad}

This tetrad serves to calculate the incoming energy flux of the
gravitational wave at infinity\cite{Kin}. It is set to be
\begin{eqnarray}
  \label{np1} l^{\mu} &=& \frac{1}{A}(1,\sqrt{AB},0,0), \\
  \label{np2} n^{\mu} &=& \frac{1}{2}(1,-\sqrt{AB},0,0), \\
  \label{np3} m^{\mu} &=& \frac{1}{\sqrt{2}r}(0,0,1,i\csc\theta),
\end{eqnarray}
of which the components queue in the order $(t,r,\theta,\varphi)$.
The corresponding spin coefficients are found to be
\begin{eqnarray}\label{npspin}
    \kappa=\sigma=\lambda=\nu=\epsilon=\pi=\tau=0,\ \ \ \ \ \ \ \ \ \ \ \ \ \ \ \ \ \\
    \rho=\frac{1}{r}\sqrt{\frac{B}{A}},\ \alpha=-\beta=\frac{\cot\theta}{2\sqrt2
    r},\ \mu=\frac{\sqrt{AB}}{2r},\
    \gamma=\frac{A_{,r}}{4}\sqrt{\frac{B}{A}}.
\end{eqnarray}
Here, taking $A=B$ recovers the Schwarzschild case.

Then we need to evaluate two contractions of the perturbation of
the Riemann tensor with the Kinnersley tetrad, which play  key
roles in calculating the energy flux at infinity and the energy
falling down the hole, and which hereafter is denoted by
$\tilde{\Psi}_0^{(1)}$ and $\tilde{\Psi}_4^{(1)}$, defined as
\begin{equation}\label{np4}
    \tilde{\Psi}_0^{(1)}=-\delta
    R_{\mu\nu\lambda\sigma}l^{\mu}m^{\nu}l^{\lambda}m^{\sigma},
\end{equation}
and
\begin{equation}\label{np100}
    \tilde{\Psi}_4^{(1)}=-\delta
    R_{\mu\nu\lambda\sigma}n^{\mu}m^{\nu*}n^{\lambda}m^{\sigma*}.
\end{equation}

To perform the contraction a new orthonormal tetrad was introduced
\begin{eqnarray}
  \label{np5}e^{\mu}_{(0)} &=& (e^{-\nu},0,0,0), \\
  \label{np6}e^{\mu}_{(1)} &=& (0,e^{-\psi},0,0), \\
  \label{np7}e^{\mu}_{(2)} &=& (0,0,e^{-\mu_2},0), \\
  \label{np8}e^{\mu}_{(3)} &=& (0,0,0,e^{-\mu_3}),
\end{eqnarray}
where  components are ordered by
$(x^0,x^1,x^2,x^3)=(t,\varphi,r,\theta)$. The components of
Kinnersley tetrad against this tetrad basis are therefore
\begin{eqnarray}
  \label{np9}l^{(\mu)} &=& (e^{-\nu},0,e^{-\nu},0), \\
  \label{np10}n^{(\mu)} &=& \frac{1}{2}(e^{\nu},0,-e^{\nu},0), \\
  \label{np11}m^{(\mu)} &=& \frac{1}{\sqrt{2}}(0,i,0,1).
\end{eqnarray}
Here the indices in the bracket indicate the components against
the basis of Eq. (\ref{np5}) through Eq. (\ref{np8}), and they run
through 0 to 3. Now we do the contraction of Eq. (\ref{np4}) and
find
\begin{eqnarray}\label{np12}
  \tilde{\Psi}_0^{(1)}=&-&ie^{-2\nu}(\delta R_{0301}
  +\delta R_{2321}+\delta R_{2301}+\delta R_{0321})\nonumber\\
    &&-\frac{1}{2}e^{-2\nu}(\delta R_{0303}+2\delta R_{0323}
    +\delta R_{2323}-\delta R_{0101}-2\delta R_{0121}-\delta
    R_{2121}),
\end{eqnarray}
as well as
\begin{eqnarray}\label{np101}
  \tilde{\Psi}_4^{(1)}=&-&\frac{i}{4}e^{2\nu}(-\delta R_{0301}
  -\delta R_{2321}+\delta R_{2301}+\delta R_{0321})\nonumber\\
    &&-\frac{1}{8}e^{2\nu}(\delta R_{0303}-2\delta R_{0323}
    +\delta R_{2323}-\delta R_{0101}+2\delta R_{0121}-\delta
    R_{2121}).
\end{eqnarray}
Here, in both expressions, the second bracket would vanish in
purely axial perturbation, whose demonstration was
raised in \S31 of \cite{Ch} and applicable to all static,
spherically symmetric metric.
For convenience, we denote the non-vanishing part,
which contains the first bracket, of Eq. (\ref{np12}) and Eq.
(\ref{np101}) to be $i\ \textrm{Im}\tilde{\Psi}_0^{(1)}$ and $i\
\textrm{Im}\tilde{\Psi}_4^{(1)}$, and the vanishing part,
$\textrm{Re}\tilde{\Psi}_0^{(1)}$ and
$\textrm{Re}\tilde{\Psi}_4^{(1)}$. The Riemann components therein
can be simplified as (\S13 of \cite{Ch}),
\begin{eqnarray}
  \label{R1}\delta R_{0301} &=& \frac{1}{2}e^{\psi-2\nu-\mu_3}Q_{03,0}
  -\frac{1}{2}e^{\psi-2\mu_2-\mu_3}Q_{23}\nu_{,r}\ , \\
  \label{R2}\delta R_{2321} &=& e^{\psi-2\mu_2-\mu_3}\left[\left(\frac{1}{r}
  -\frac{1}{2}\mu_{2,r}\right)Q_{23}+\frac{1}{2}Q_{23,r}\right], \\
  \label{R3}\delta R_{2301} &=& \frac{1}{2}e^{\psi-\nu-\mu_2-\mu_3}\left[Q_{03,2}-Q_{02,3}
  +Q_{03}\left(\frac{1}{2}+\mu_{2,r}\right)-Q_{02}\cot\theta\right], \\
  \label{R4}\delta R_{0321} &=&-\frac{1}{2}e^{\psi-\nu-\mu_2-\mu_3}
  \left[Q_{20,3}+Q_{20}\cot\theta-Q_{23,0}+Q_{30}\left(\frac{1}{r}
  -\nu_{,r}\right)\right].
\end{eqnarray}
Inserting these equations into Eq. (\ref{np12}) and Eq.
(\ref{np101}), we have
\begin{eqnarray}\label{np20}
  -\frac{e^{2\nu+2\mu_2}}{\sin\theta}\textrm{Im}\tilde{\Psi}_0^{(1)}
  = \frac{Q_{23}}{r}+\frac{1}{2}Q_{23,r}+\frac{1}{2}e^{-\nu+\mu_2}Q_{23,0}
  -\frac{1}{2}(\mu_{2,r}+\nu_{,r})Q_{23}\nonumber\\
  +\frac{1}{2}e^{-\nu+\mu_2}\left[Q_{03,r}+\left(\frac{2}{r}-\nu_{,r}+\mu_{2,r}\right)
  Q_{03}\right]+\frac{1}{2}e^{-2\nu+2\mu_2}Q_{03,0},
\end{eqnarray}
and
\begin{eqnarray}\label{np102}
  -\frac{4e^{-2\nu+2\mu_2}}{\sin\theta}\textrm{Im}\tilde{\Psi}_4^{(1)}
  = -\frac{Q_{23}}{r}-\frac{1}{2}Q_{23,r}+\frac{1}{2}e^{-\nu+\mu_2}Q_{23,0}
  +\frac{1}{2}(\mu_{2,r}+\nu_{,r})Q_{23}\nonumber\\
  +\frac{1}{2}e^{-\nu+\mu_2}\left[Q_{03,r}+\left(\frac{2}{r}-\nu_{,r}+\mu_{2,r}\right)
  Q_{03}\right]-\frac{1}{2}e^{-2\nu+2\mu_2}Q_{03,0}.
\end{eqnarray}
With the aid of Eq. (\ref{g5}) through Eq. (\ref{g8}), the two
contractions can be simplified to be
\begin{eqnarray}\label{scR}
  2i\omega\  \textrm{Im}\tilde{\Psi}_0^{(1)}=\sum_{l=2}^{\infty}
  \left\{\left(\frac{2\sqrt{AB}}{r}-\frac{A'B+AB'}{2\sqrt{AB}}-2i\omega\right)
  \Lambda_{(-)}Z_l+\left[\frac{i\omega(AB'-A'B)}{2\sqrt{AB}}+V_{g,l}\right]Z_l\right\}
  \frac{C^{-3/2}_{l+2}(\cos\theta)}{rA^2\sin^2\theta},
\end{eqnarray}
and
\begin{eqnarray}\label{np103}
  8i\omega\  \textrm{Im}\tilde{\Psi}_4^{(1)}=\sum_{l=2}^{\infty}
  \left\{\left(\frac{2\sqrt{AB}}{r}-\frac{A'B+AB'}{2\sqrt{AB}}-2i\omega\right)
  \Lambda_{(+)}Z_l+\left[\frac{i\omega(A'B-AB')}{2\sqrt{AB}}+V_{g,l}\right]Z_l\right\}
  \frac{C^{-3/2}_{l+2}(\cos\theta)}{r\sin^2\theta},
\end{eqnarray}
where $V_{g,l}$ is the radial effective potential defined by Eq.
(\ref{g11}) and $\Lambda_{(\pm)}=d/dr_*\pm i\omega$. It is easy to
check that equating $A(r)$ and $B(r)$ would recover the result in the
Schwarzschild case.

\subsubsection{Formalism in the Hawking-Hartle Tetrad}

To calculate the gravitational wave energy flux on the horizon,
the Kinnersley tetrad is no longer applicable because of their
singular behavior at the horizon. We use the tetrad introduced by
Hartle and Hawking instead\cite{HH}, which represents a physical
observer crossing the event horizon. It eliminates the singular
behavior by imposing on the Kinnersley tetrad a rotation of the
third class (\cite{Ch}, \S8). In our problem the rotation
parameter is $\Lambda=2/A$.  The tetrad is normalized so  that on
the basis of coordinate $(v=t+r_*,r,\theta,\varphi)$,
$(l^{HH})^v=1$. This results in the Hawking-Hartle basis as
follows:
\begin{eqnarray}
  \label{hh1}(l^{HH})^{\mu} &=& 1/2(1,\sqrt{AB},0,0), \\
  \label{hh2}(n^{HH})^{\mu} &=& (1/A,-\sqrt{B/A},0,0),
\end{eqnarray}
and $m^{\mu}$ is unchanged. The components are arranged in the
order of $(t,r,\theta,\varphi)$. Changing to the basis of
$(\partial_v,\partial_r,\partial_{\theta},\partial_{\varphi})$, we
have
\begin{eqnarray}
  \label{hh3}(l^{HH})^{\mu} &=& (1,\sqrt{AB}/2,0,0), \\
  \label{hh4}(n^{HH})^{\mu} &=& (0,-\sqrt{B/A},0,0),
\end{eqnarray}
again with $m^{\mu}$ unchanged. Thus we obtain the well-behaved
tetrad on the horizon, which is equivalent to a physical observer
advancing along the direction $l^{HH}$. Hereafter we perform
calculations in this Hawking-Hartle tetrad. We can derive new spin
coefficients using their transformation rules under the third
class of tetrad rotation (\cite{Ch}, \S8), the one of importance
is
\begin{equation}\label{hh5}
    \epsilon^{HH}=-\frac{1}{2}\Lambda^{-2}D\Lambda
    =\frac{1}{4}\sqrt{\frac{B}{A}}\frac{dA}{dr},
\end{equation}
where $D$ is the differential operator corresponding to the
tangent vector $l^{\mu}$. We define
\begin{equation}\label{hh6}
    \epsilon_0=\epsilon^{HH}_{r_H}=\left.\frac{1}{4}
    \sqrt{\frac{B}{A}}\frac{dA}{dr}\right|_{r_H}
    =\frac{\sqrt{A_1B_1}}{4r_H}.
\end{equation}
$2\epsilon_0$ is the surface gravity of the black hole (\cite{Ch},
\S8). We also need  to know that $\kappa$ vanishes globally,
ensuring the integral curves of $l^{HH}$ to be null geodesics, and
that $\rho^{HH}$ vanishes at the horizon, since
\begin{equation}\label{hh7}
    \rho^{HH}=\Lambda^{-1}\rho=\frac{\sqrt{AB}}{2r}\longrightarrow 0,
    \ \ (r\rightarrow r_H).
\end{equation}
Moreover, we present here two relations needed for calculating the
energy falling into the hole. They are obtained with the
perturbation effect, where the Hawking-Hartle tetrad is also
perturbed so that $l^{HH}$ is the generator of a congruence of
null geodesics, and  $\kappa$ always stays zero. We start from two
Ricci identities under any Newmann-Penrose basis, which are
\begin{eqnarray}
  D\sigma-\delta\kappa &=& \sigma(3\epsilon-\epsilon^*+\rho+\rho^*)
  +\kappa(\pi^*-\tau-3\beta-\alpha^*)+\Psi_0, \\
  D\rho-\delta^*\kappa &=& (\rho^2+|\sigma|^2)+\rho(\epsilon+\epsilon^*)
  -\kappa^*\tau-\kappa(3\alpha+\beta^*-\pi)+\Phi_{00}
\end{eqnarray}
(\cite{Ch}, \S8). Here $\Psi_0$ and $\Phi_{00}$ are projections of
the Weyl tensor and the Ricci tensor upon the tetrad. Linearizing
these two and with the consequence from $\delta R_{\mu\nu}=0$ that
$\Phi^{(1)}_{00}=0$, we find, under Hawking-Hartle tetrad,
\begin{eqnarray}
  \label{hh11}D^{HH}_{r_H}\sigma^{HH(1)}_{r_H} &=&
  2\epsilon_0\sigma^{HH(1)}_{r_H}+\left.\Psi^{HH(1)}_0\right|_{r_H}\ , \\
  \label{hh12}D^{HH}_{r_H}\rho^{HH(1)} &=& 2\epsilon_0\rho^{HH(1)}.
\end{eqnarray}

\subsection{Gravitational Wave Energy Absorption}

\subsubsection{Incoming Energy Flux at Infinity}
The general expression for the incoming energy flux is
\cite{eFlow, early9, Ch}
\begin{equation}\label{i1}
    \frac{d^2E^{(inc)}}{dtd\Omega}=\frac{1}{64\pi\omega^2}
    \lim_{r\rightarrow\infty}
    r^2|\tilde{\Psi}_0^{(1)}|^2,
\end{equation}
where $\tilde{\Psi}_0^{(1)}$ is defined in Eq. (\ref{np12}).
Considering the axial perturbation, we can replace
$\tilde{\Psi}_0^{(1)}$ by $\textrm{Im}\tilde{\Psi}_0^{(1)}$.
Taking $r\rightarrow\infty$ in Eq. (\ref{scR}) and taking into
account only the incoming wave, we have
\begin{equation}\label{i2}
    \textrm{Im}\tilde{\Psi}_0^{(1)}\longrightarrow\sum_{l=2}^{\infty}
    \left(\Lambda_{(-)}Z_l^{(inc)}\right)\frac{C_{l+2}^{-3/2}(\cos\theta)}{r\sin^2\theta}
    \ ,\ \ (r\rightarrow\infty).
\end{equation}
With the boundary condition at infinity Eq. (\ref{g21}), we have
\begin{equation}\label{i3}
    \Lambda_{(-)}Z_l^{(inc)}\longrightarrow(-1)^{l+1}i\omega(2l+1)e^{-i\omega r_*}/6,\
    \ (r\rightarrow\infty).
\end{equation}
Inserting Eq. (\ref{i2}), Eq. (\ref{i3}) into Eq. (\ref{i1}) and
integrating the flux over all directions, we obtain the total
incoming energy per unit time at infinity
\begin{equation}\label{i4}
    \Phi^{(inc)}=\int_{4\pi}{\frac{d^2E^{(inc)}}{dtd\Omega}\ d\Omega}
    =\frac{\omega^2}{2304\pi}\sum_{l=2}^{\infty}\int_{4\pi}
    {\left[\frac{C_{l+2}^{-3/2}(\cos\theta)}{\sin^2\theta}\right]^2d\Omega}
    =\frac{\omega^2}{192}.
\end{equation}
The last equality was obtained by using the orthogonality of the
Gegenbauer polynomials  \cite{ToI}
\begin{equation}\label{GegOrth}
    \int_{-1}^1\frac{C_{l+2}^{-3/2}C_{l'+2}^{-3/2}}
    {(1-x^2)^2}dx=\frac{18\ \delta_{ll'}}{(2l+1)(l+2)(l+1)l(l-1)},
\end{equation}
and the summation
\begin{equation}\label{i5}
    \sum_{l=2}^{\infty}\frac{2l+1}{(l+2)(l+1)l(l-1)}=\frac{1}{3}.
\end{equation}

\subsubsection{Energy Falling into the Hole}

Calculating the gravitational wave energy absorption into the
black hole is more complicated than that in the scalar case. We
will follow Teukolsky's treatment \cite{early9} to deal with this
problem.

With Eq. (\ref{hh12}), following the procedure
presented in \cite{Ch}, \S98, we have the change in the area of
the event horizon in the process of perturbation
\begin{equation}\label{i7}
    \frac{d^2\Sigma}{dvd\Omega}=\frac{r_H^2}{\epsilon_0}|\sigma_{r_H}^{HH(1)}|^2,
\end{equation}
where $\Sigma$ denotes the area of event horizon, and the spin
coefficient $\epsilon_0$ is presented in  Eq. (\ref{hh6}). The
first law of black hole thermodynamics relates the change of the
horizon area to the change of the black hole internal energy
\begin{equation}\label{i8}
    dE=\frac{\bar{\kappa}}{8\pi} d\Sigma,
\end{equation}
where $\bar{\kappa}=2\epsilon_0$ is the surface gravity. Thus
\begin{equation}\label{i9}
    \frac{d^2\Sigma}{dvd\Omega}=\frac{4\pi}
    {\epsilon_0}\frac{d^2E^{(abs)}}{dvd\Omega}.
\end{equation}
Combining Eq. (\ref{i8}) and Eq. (\ref{i9}), we have
\begin{equation}\label{i10}
    \frac{d^2E^{(abs)}}{dvd\Omega}=\frac{r_H^2}{4\pi}
    |\sigma_{r_H}^{HH(1)}|^2.
\end{equation}
The spin coefficient $\sigma_{r_H}^{HH(1)}$ is evaluated in Eq.
(\ref{hh11}), where
\begin{equation}\label{i11}
    D_{r_H}^{HH}=\partial_v=\partial_t\rightarrow-i\omega,
\end{equation}
and consequently we get
\begin{equation}\label{i12}
    \sigma_{r_H}^{HH(1)}=-\frac{\Psi^{HH(1)}_0|_{r_H}}{i\omega+2\epsilon_0}.
\end{equation}
Referring to the transformation of Weyl scalars under the rotation
of the third class (\cite{Ch}, \S8), we have
\begin{equation}\label{i13}
    \Psi^{HH(1)}_0=A^2\Psi_0^{(1)}/4.
\end{equation}
Combining Eq. (\ref{i13}), Eq. (\ref{i12}) and Eq. (\ref{i10}), we
arrive at
\begin{equation}\label{i14}
    \frac{d^2E^{(abs)}}{dvd\Omega}=\frac{r_H^2}{16\pi}
    \frac{|A^2\Psi^{(1)}_0|^2_{r_H}}{4\omega^2+A_1B_1r_H^{-2}}.
\end{equation}
Now we need to evaluate $\Psi^{(1)}_0=-\delta
C_{\mu\nu\rho\lambda}l^{\mu}m^{\nu}l^{\rho}m^{\lambda}$,
\begin{eqnarray}\label{i15}
  \Psi_0^{(1)}=&-&ie^{-2\nu}(\delta C_{0301}
  +\delta C_{2321}+\delta C_{2301}+\delta C_{0321})\nonumber\\
    &&-\frac{1}{2}e^{-2\nu}(\delta C_{0303}+2\delta C_{0323}
    +\delta C_{2323}-\delta C_{0101}-2\delta C_{0121}-\delta
    C_{2121}).
\end{eqnarray}
The Weyl tensors are defined as
\begin{equation}\label{Weyl}
    C_{\mu\nu\rho\lambda}=R_{\mu\nu\rho\lambda}
    -\frac{1}{2}(g_{\mu\rho}R_{\lambda\nu}-g_{\mu\lambda}
    R_{\rho\nu}+g_{\nu\rho}R_{\lambda\mu}-g_{\nu\lambda}R_{\rho\mu})
    +\frac{1}{3}R(g_{\mu\rho}g_{\lambda\nu}-g_{\mu\lambda}g_{\rho\nu}).
\end{equation}
Three facts can help greatly simplify the evaluation, which are,
(i) only the diagonal components of the Ricci tensor is non-zero
for a metric like Eq. (\ref{e3}), (ii) according to the
perturbation equation, $\delta R_{\mu\nu}=0$, and (iii) only
$\delta g_{1\mu}\neq0$, $(\mu=0,2,3)$. The evaluation shows that
the components of the perturbed Weyl tensors in $\Psi^{(1)}_0$ all
coincide with those of the Riemann's, except for $\delta C_{0301}$
and $\delta C_{2321}$, which are
\begin{eqnarray}
\label{i16}
    \delta C_{0301}=\delta R_{0301}+\frac{1}{2}R_{00}\delta
    g_{31}+\frac{1}{3}Rg_{00}\delta g_{31},\\
\label{i17}
    \delta C_{2321}=\delta R_{2321}+\frac{1}{2}R_{22}\delta
    g_{31}+\frac{1}{3}Rg_{22}\delta g_{31}.
\end{eqnarray}
Therefore
\begin{equation}\label{i18}
    \Psi^{(1)}_0=\tilde{\Psi}^{(1)}_0+\frac{ie^{-2\nu}}{2}(R_{00}+R_{22})
    \delta g_{31}+\frac{ie^{-2\nu}}{3}R(g_{00}+g_{22})\delta g_{31}.
\end{equation}
This is further simplified by the fact that we carry out the
contraction under the orthonormal basis, Eq. (\ref{np5}) through
Eq. (\ref{np8}), where for a spherical metric as Eq. (\ref{e3})
describes, there holds $g_{00}=-g_{22}=1$, implying that
$(g_{00}+g_{22})\delta g_{31}\equiv 0$. And also, we have the
extended result from the calculation of the Schwarzschild model
\begin{eqnarray}\label{i1000}
    &&R_{00}=-\frac{BA''}{2A}-\frac{A'B'}{4A}-\frac{BA'}{rA}+\frac{B}{4A^2}(A')^2;
    \nonumber\\
    &&R_{22}=\frac{BA''}{2A}+\frac{A'B'}{4A}+\frac{B'}{r}-\frac{B}{4A^2}(A')^2,
\end{eqnarray}
which gives that
\begin{equation}\label{i1001}
    R_{00}+R_{22}=\frac{B}{r}(\frac{B'}{B}-\frac{A'}{A})\sim
    O(r/r_H-1),\ \ (r\rightarrow r_H+0).
\end{equation}
That is, the term $R_{00}+R_{22}$ is of the first order small
quantity as $r$ tends to the horizon. On the other hand, it is
easy to check that near the horizon, we have $q_{3,2}\sim Q
(r/r_H-1)^{-1}\sim(r/r_H-1)^{-i(\omega/\sqrt{A_1B_1})-1}$, so that
$\delta g_{31}\sim q_{3} \sim
(r/r_H-1)^{-i(\omega/\sqrt{A_1B_1})}$, and hence we have
$(R_{00}+R_{22})\delta g_{31}\rightarrow 0$, as $r\rightarrow
r_H+0$. Therefore in our context
\begin{equation}\label{i19}
    \Psi_0^{(1)}=i\ \textrm{Im}\tilde{\Psi}_0^{(1)}.
\end{equation}
Hence
\begin{equation}\label{i20}
    \frac{d^2E^{(abs)}}{dvd\Omega}=\frac{r_H^2}{16\pi}
    \frac{|A^2\textrm{Im}\tilde{\Psi}^{(1)}_0|^2_{r_H}}{4\omega^2+A_1B_1r_H^{-2}}.
\end{equation}
In the vicinity of the event horizon, $r\rightarrow r_H+0$,
applying the boundary condition  Eq. (\ref{g26}), we get
\begin{eqnarray}\label{i21}
    &&A^2\textrm{Im}\tilde{\Psi}^{(1)}_0\rightarrow\sum_{l=2}^{\infty}
    \left[\left(\sqrt{A_1B_1}r_H^{-1}-2i\omega\right)\Lambda_{(-)}Z_l\right]
    \frac{C_{l+2}^{-3/2}(\cos\theta)}{2i\omega r_H
    \sin^2\theta}\ \ \ \ \ \ \ \nonumber\\
    &&\rightarrow-\sum_{l=2}^{\infty}
    \left[\left(\sqrt{A_1B_1}r_H^{-1}-2i\omega\right)
    p_l(r/r_H-1)^{-i\omega r_H/\sqrt{A_1B_1}}\right]
    \frac{C_{l+2}^{-3/2}(\cos\theta)}{r_H\sin^2\theta}.
\end{eqnarray}
Inserting this expression into Eq. (\ref{i20}) and performing the
integration over all directions, we have the total absorbed energy
per unit time
\begin{eqnarray}\label{i21}
    \Phi^{(abs)}&=&\int_{4\pi}{\frac{d^2E^{(abs)}}{dvd\Omega}\ d\Omega}
    =\frac{|p_l|^2}{16\pi}\int_{4\pi}
    {\left[\frac{C_{l+2}^{-3/2}(\cos\theta)}{\sin^2\theta}\right]^2d\Omega}.
\end{eqnarray}
Employing the orthogonality of the Gagenbauer polynomials, Eq.
(138) becomes
\begin{equation}\label{i22}
    \Phi^{(abs)}=\frac{9}{4}\sum_{l=2}^{\infty}\frac{|p_l|^2}{2l+1}
    \frac{(l-2)!}{(l+2)!}.
\end{equation}

\subsubsection{The Absorption Cross Section}

We define the absorption cross section as the ratio of the total
absorbed energy to the total incoming energy at infinity.
Averaging the total incident energy over the event
horizon, we have the incident energy flux in the analogous sense,
that
\begin{equation}\label{i23}
    j^{(inc)}=\frac{\Phi^{(inc)}}{4\pi r_H^2}
    =\frac{\omega^2}{1536\pi r_H^2}.
\end{equation}
The total absorption cross section reads
\begin{equation}\label{GTotSec}
    \sigma_g^{abs}=\frac{\Phi^{(abs)}}{j^{(inc)}}=\frac{1728\pi
    r_H^2}{\omega^2}\sum_{l=2}^{\infty}\frac{|p_l|^2}{2l+1}
    \frac{(l-2)!}{(l+2)!},
\end{equation}
The partial absorption cross
section has the form
\begin{equation}\label{GParSec}
    \sigma_{g,l}^{abs}=\frac{1728\pi
    r_H^2}{\omega^2}\frac{|p_l|^2}{2l+1}
    \frac{(l-2)!}{(l+2)!}.
\end{equation}
Now we proceed to evaluate $p_l$, which is necessary to obtain the

By means of the Wronskian of the
radial perturbation equation Eq. (\ref{g10}), for any two
solutions $Z_l^{(1)}$ and $Z_l^{(2)}$, their Wronskian is
\begin{equation}\label{i51}
    W[Z_l^{(1)},Z_l^{(2)}]=\frac{C_l^{(12)}}{\sqrt{AB}},
\end{equation}
where $C_l^{(12)}$ is a complex number depending on the explicit
form of $Z_l^{(1)}$ and $Z_l^{(2)}$. Applying this to the
asymptotic solution both at infinity, as is shown in Eq.
(\ref{g21}), and at the horizon, as is shown in Eq. (\ref{g26}),
we have, for the region at infinity,
\begin{equation}\label{i52}
    W[Z_l^*, Z_l]=-\frac{i\omega^3(2l+1)^2}{72\sqrt{AB}}
    \left(1-|S^g_l|^2\right)=-\frac{i\omega^3(2l+1)^2}{72\sqrt{AB}}
    \Gamma^g_l,
\end{equation}
and for the region approaching the horizon,
\begin{equation}\label{i53}
    W[Z_l^*, Z_l]=-\frac{2i\omega}{\sqrt{AB}}|p_l|^2.
\end{equation}
Equating the above two equations gives
\begin{equation}\label{i54}
    |p_l|^2=\frac{\omega^2(2l+1)^2}{144}\Gamma^g_l,
\end{equation}
where
\begin{equation}\label{i50}
    \Gamma^g_l=1-|S^g_l|^2,
\end{equation}

Inserting this $|p_l|^2$ into the expression of partial absorption
cross sections, Eq. (\ref{GParSec}), we get
\begin{equation}\label{GOpT}
    \sigma^{abs}_{g,l}=12\pi r_H^2
    (2l+1)\frac{(l-2)!}{(l+2)!}\Gamma^g_l,
\end{equation}
which is analogous to Eq. (\ref{SOT}) in the scalar case. In the
high energy limit, the wave can easily penetrate the potential
barrier and be absorbed by the black hole,  $\Gamma_l^g\rightarrow
1$. Thus summing up both sides of Eq. (\ref{GOpT}) and applying
the summation formula Eq. (\ref{i5}), we get
\begin{equation}\label{i56}
    \sigma^{abs}_g\longrightarrow4\pi r_H^2, \ \
    (\omega\rightarrow\infty).
\end{equation}

\subsection{Conservation of Energy}
Here we will justify the formalism for the energy absorption rate
in \S III-D-2. In the far away region, we can write Eq(70) in the
form
\begin{equation}\label{CEF1}
    Q\sim e^{-i\omega t}\sum_{l=2}^{\infty}r\left(\mathcal{I}_le^{-i\omega r_*}+\mathcal{R}_le^{i\omega
    r_*}\right)C_{l+2}^{-3/2}(\cos\theta), \ \ r\rightarrow\infty.
\end{equation}
Near the horizon we can follow Eq. (\ref{g27}) and express
\begin{equation}\label{CEF2}
    Q\sim e^{-i\omega t}r_H(r/r_H-1)^{-i\omega r_H/\sqrt{A_1B_1}}
    \sum_{l=2}^{\infty}\mathcal{T}_lC^{-3/2}_{l+2}(\cos\theta),\ \
    r\rightarrow r_H+0,
\end{equation}
where we have replaced $p_l$ with $\mathcal{T}_l$ to avoid
ambiguity. Our goal is to show that the difference between the
incoming energy and the reflected energy by the potential barrier
at infinity is all absorbed by the black hole
\begin{equation}\label{CEF3}
    \frac{dE^{(abs)}}{dt}=\frac{dE^{(inc)}}{dt}-\frac{dE^{(ref)}}{dt}.
\end{equation}
For the incoming energy per unit time, we adopt  Eq.
(\ref{i1}), and get
\begin{equation}\label{CEF4}
    \frac{dE^{(inc)}}{dt}=\sum_{l=2}^{\infty}\frac{|\mathcal{I}_l|^2}{16\pi}
    \Xi_l,
\end{equation}
where we have employed the orthogonality of the Gegengauer
polynomials and we let
\begin{equation}\label{CEF00}
    \int_{4\pi}\left[\frac{C_{l+2}^{-3/2}(\cos\theta)}{\sin^2\theta}\right]^2d\Omega
    =\Xi_l
\end{equation}
for simplicity. The reflected energy can be obtained similarly.
Referring to ref. \cite{Ch}, \S98, the out flowing energy at
infinity is
\begin{equation}\label{CEF5}
    \frac{d^2E^{(ref)}}{dtd\Omega}=\frac{1}{16\pi\omega^2}
    \lim_{r\rightarrow \infty}|\tilde{\Psi}^{(1)}_4|^2.
\end{equation}
Inserting  Eq. (\ref{np103}) into it and using Eq. (\ref{CEF1}),
we have
\begin{equation}\label{cef6}
    \frac{dE^{(ref)}}{dt}=\sum_{l=2}^{\infty}\frac{|\mathcal{R}_l|^2}{16\pi}
    \Xi_l.
\end{equation}
The absorbed energy per unit time can be got by referring to the
calculation in \S III-D-2 and using  Eq. (\ref{CEF2})
\begin{equation}\label{CEF7}
    \frac{dE^{(abs)}}{dt}=\sum_{l=2}^{\infty} \frac{|\mathcal{T}_l|^2}{16\pi}
    \Xi_l.
\end{equation}
To complete the demonstration, we need to establish the relation
among $\mathcal{I}_l$, $\mathcal{R}_l$ and $\mathcal{T}_l$. We do
this by means of Wronskians of the radial equation Eq.
(\ref{g10}). Referring to the last section where we have
calculated the Wronskian, and by virtue of Eq. (\ref{CEF1}) and Eq. (\ref{CEF2}), we have the Wronskian at
infinity that
\begin{equation}\label{CEF8}
    W[Z^*_l,Z_l]=2i\omega\left(-|\mathcal{I}_l|^2+|\mathcal{R}_l|^2\right)(AB)^{-1/2},
\end{equation}
and also the Wronskian at the horizon that
\begin{equation}\label{CEF9}
    W[Z^*_l,Z_l]=-2i\omega|\mathcal{T}_l|^2(AB)^{-1/2}.
\end{equation}
Equating the above two, we have the relation among
$\mathcal{I}_l$, $\mathcal{R}_l$ and $\mathcal{T}_l$ that
\begin{equation}\label{CEF10}
    |\mathcal{T}_l|^2=|\mathcal{I}_l|^2-|\mathcal{R}_l|^2.
\end{equation}
Now we can relate Eq. (\ref{CEF4}), Eq. (\ref{cef6}) and Eq.
(\ref{CEF7}) by using Eq. (\ref{CEF3}), thus completing the
demonstration. This demonstration shows that the application of
Hawking-Hartle tetrad in our above discussions is well grounded on
the law of energy conservation.

\subsection{Numerical Study for Gravitational Perturbation}

\begin{figure}
  \includegraphics[width=9cm]{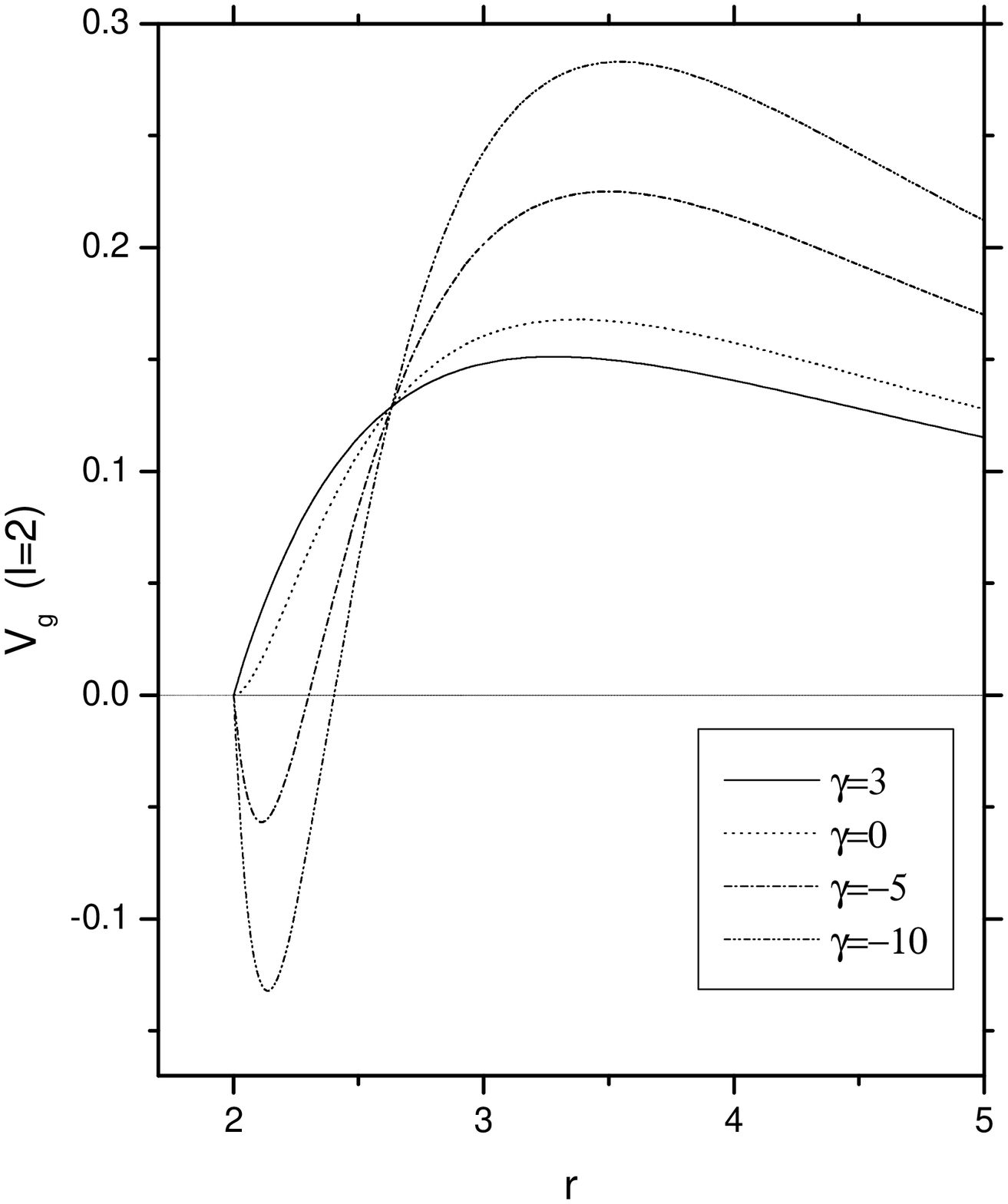}\includegraphics[width=9cm]{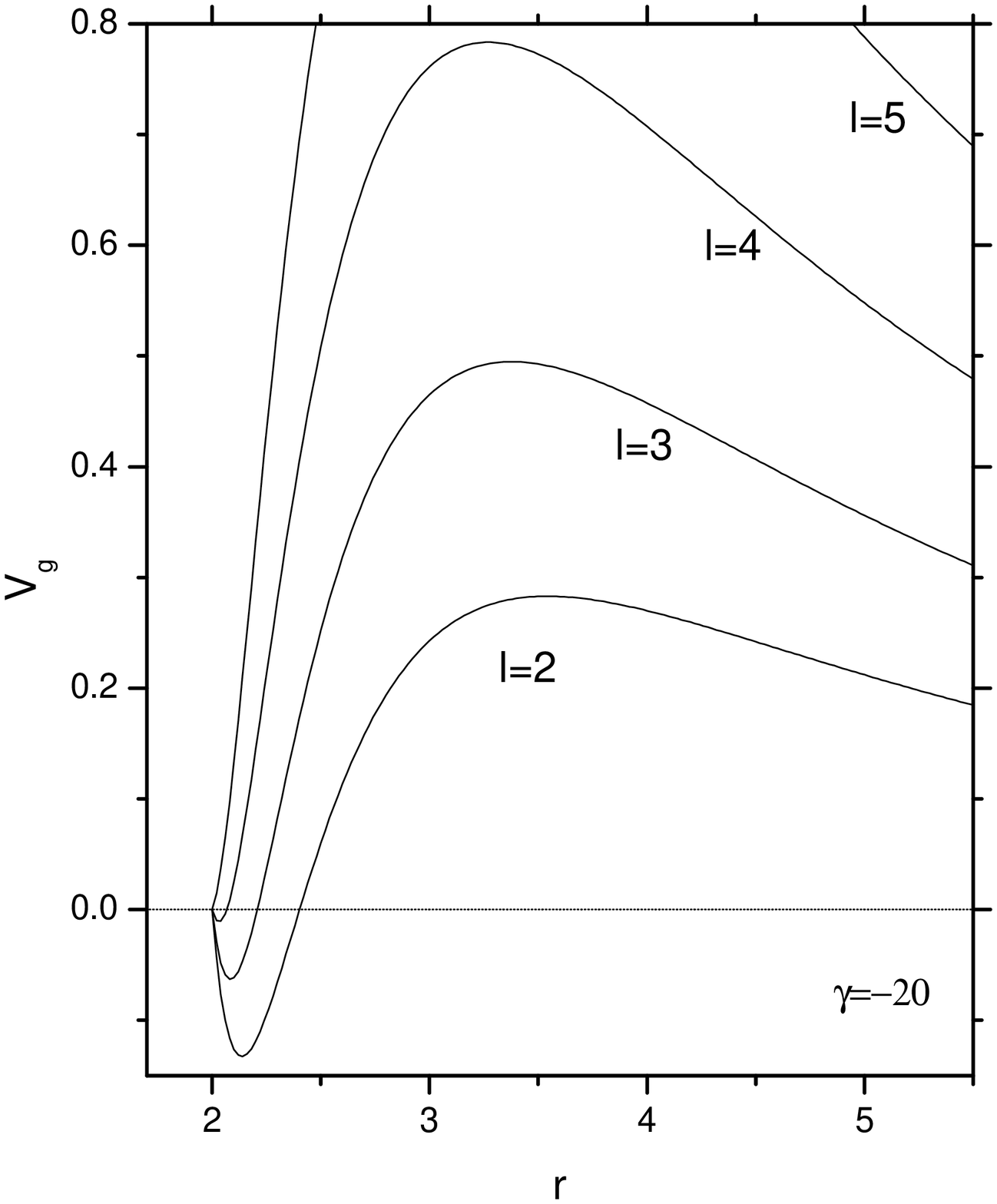}\\
  \caption{The effective potentials of the gravitational perturbation for different
  $\gamma$ and $l$.}\label{VgGm}
\end{figure}

\begin{figure}
  \includegraphics[width=9cm]{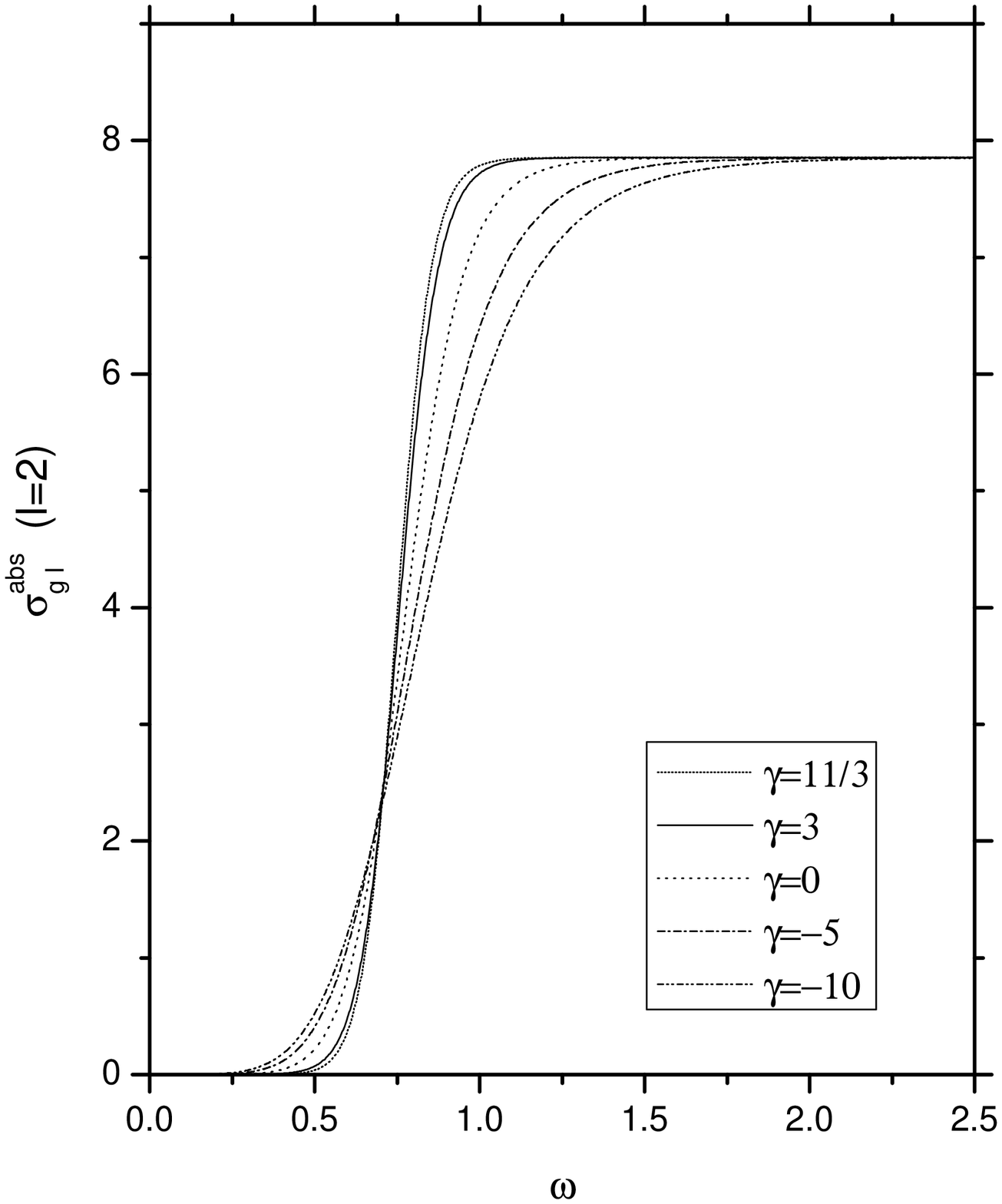}\includegraphics[width=9cm]{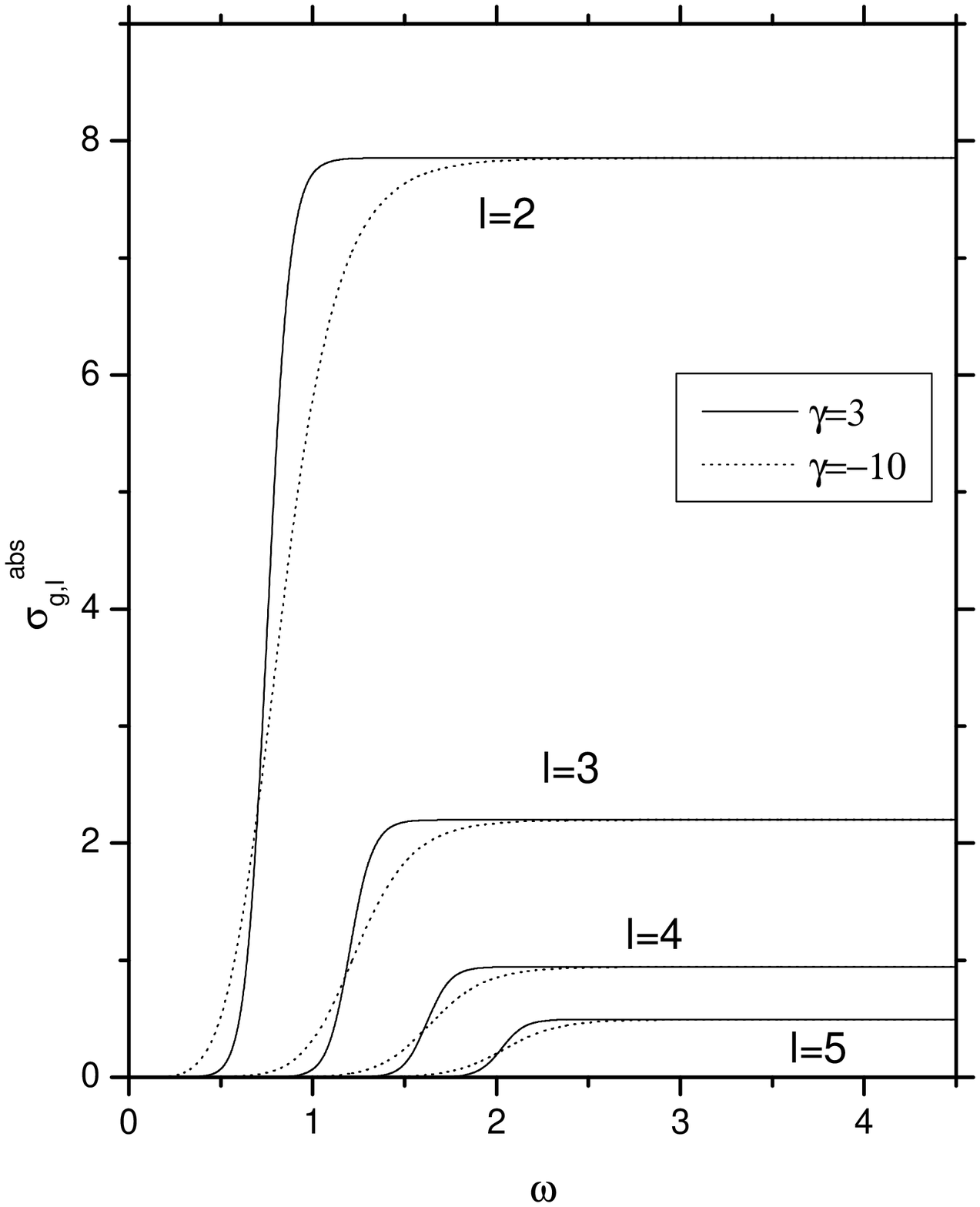}\\
  \caption{Partial energy absorption of the gravitational wave for different $\gamma, l$.}
  \label{PGl2}
\end{figure}

\begin{figure}
  \includegraphics[width=12cm]{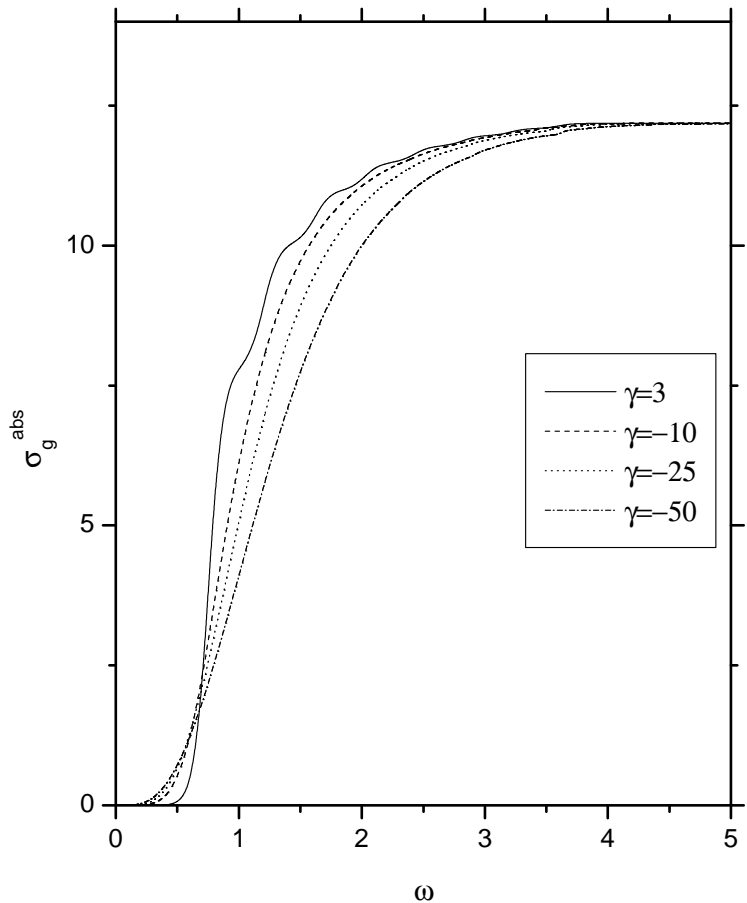}\\
  \caption{Total energy absorption of the gravitational wave.}\label{totGrav}
\end{figure}

In solving the radial equation Eq. (\ref{g10}), we
choose a particular solution $\phi_l(r)$ to be normalized as
\begin{equation}\label{gnm10000}
    \phi_l(r)=p_l^{-1}Z_l(r).
\end{equation}
In the remote region, it consists of the outgoing and ingoing
waves, which is
\begin{equation}\label{gnm3}
    \phi_l(r)=n_l^{(-)}\mathcal{N}_{l(+)}+n_l^{(+)}\mathcal{N}_{l(-)},
\end{equation}
where
\begin{equation}\label{gnm2}
    \mathcal{N}_{l(\pm)}(r)=e^{\mp i\omega
    r_*}\sum_{n=0}^{\infty}\kappa^{(\pm)}_{l,n}\left(\frac{r_H}{r}\right)^n,
\end{equation}
and the combination coefficients $n_l^{(\pm)}$ are called jost
functions. To obtain the jost functions numerically, we calculate
their Wronskian, by applying Eq. (\ref{i51}) we arrive at
\begin{equation}\label{gnm4}
    W[\mathcal{N}_{l(+)},\mathcal{N}_{l(-)}]=\frac{2i\omega
    }{\sqrt{AB}}.
\end{equation}
The jost functions read
\begin{equation}\label{gnm5}
    n_l^{(\pm)}=\mp\frac{\sqrt{AB}}{2i\omega} W[\phi_l, \mathcal{N}_{l(\pm)}].
\end{equation}
Comparing Eq. (\ref{gnm3}) with Eq. (\ref{g21}), we have
\begin{eqnarray}\label{gnm6}
    p_l=(-1)^l\frac{(2l+1)\omega}{12\ n_l^{(-)}},\\
\label{gnm7}    S_l^g=(-1)^{l+1}\frac{n_l^{(+)}}{n_l^{(-)}}.\ \ \
\end{eqnarray}
They are key quantities for determining the absorption cross
section.

In practice, we insert Eq. (\ref{gnm6}) into Eq. (\ref{GTotSec})
and Eq. (\ref{GParSec}), and have
\begin{equation}\label{gnm8}
    \sigma_{g,l}^{abs}=12\pi
    r_H^2\frac{(l-2)!}{(l+2)!}\frac{2l+1}{|n_l^{(-)}|^2},
\end{equation}
and
\begin{equation}\label{gnm7}
    \sigma_g^{abs}=12\pi
    r_H^2\sum_{l=2}^{\infty}
    \frac{(l-2)!}{(l+2)!}\frac{2l+1}{|n_l^{(-)}|^2}.
\end{equation}

Now we can do the numerical calculation. Substituting the CFM
black hole metric, for the gravitational perturbation, the
effective potential has the form
\begin{equation}\label{potg}
    V_{g,l}(r)=\left(1-\frac{2M}{r}\right)
    \left[\frac{l(l+1)}{r^2}-\frac{6M}{r^3}-
    \frac{M(\gamma-3)(5r^2-20Mr+18M^2)}{r^3(2r-3M)^2}\right].
\end{equation}
Choosing $M=1$, we display the behavior of the potential for the
axial perturbation in Fig.4. The effective potential is not
positive definite. For enough negative value of $\gamma$, a
negative peak will show up out of the horizon. The negative
potential well becomes deeper when $\gamma$ becomes more negative
for fixed $l$. For chosen $\gamma$, the negative potential well
appears for small $l$. Compared with the scalar case, in the
gravitational perturbation, the negative wells appear before the
potential barriers. It was argued that negative potential may
result in the amplification of the perturbation out of the black
hole and cause the spacetime to be unstable \cite{wang01}. However
in the QNM study of the CFM braneworld black hole \cite{eabdalla},
it was shown that even for very negative value of $\gamma$, the
perturbative dynamics out of the black hole is always stable. The
negative potential well loses the competition with the positive
potential barrier and the perturbative dynamics is still dominated
by the positive potential barrier.

In the study of the absorption, the potential well tends to
enhance the absorption while the barrier tends to decrease it. The
result on the partial absorption cross section is shown in Fig.5.
We find that as the case observed in the QNM study
\cite{eabdalla}, the potential barrier wins the competition with
the well and dominates the contribution to the absorption rate.
Fixing $l$, we observe that for more negative $\gamma$, less
gravitational wave energy is absorbed by the CFM braneworld black
hole. And for chosen $\gamma$, the absorption rate decreases with
the increase of $l$ due to higher potential barrier. In the low
$\omega$ region, the absorption is enhanced for smaller values of
$\gamma$, which is the effect of the negative potential well.

In the high energy limit, the partial absorption cross sections of
the gravitational wave flatten out which is different from the
case in the scalar wave. This is due to the difference in Eq.
(\ref{SOT}) and Eq. (\ref{GOpT}), where in the scalar case the
partial section is proportional to $\omega^{-2}$ in the high
frequency regime, while in the gravitational wave case, the
partial section is proportional to $\omega^0$ in the high
frequency regime. Deep reason causing this difference lies in the
different expansions of the plane wave. In the scalar case, the
plane wave is expanded in terms of Legendre polynomials into
spherical waves with a factor $(\omega r)^{-1}e^{\pm i\omega r}$
(Eq. [\ref{f6}]), while in the gravitational case, it is expanded
in Gegenbauer polynomials into the wave with a factor $(\omega r)
e^{\pm i\omega r}$ (Eq. [\ref{g15}]).

The result on the total absorption cross section of the
gravitational wave is shown in Fig.6. The total absorption cross
section, which is the physically observable quantity, behaves in a
similar manner to that in the scalar case. This shows that the
difference in the gravitational partial absorption cross section
from that of the scalar case is more mathematical than physical.


\section{Conclusion and Discussion}
In this paper, we have studied the energy absorption problem of
the scalar wave as well as the axial gravitational wave in the
background of the CFM brane-world black hole. The CFM black hole
is spherical and has only one event horizon at $r_H=2M$. When the
parameter $\gamma=3$, the CFM braneworld black hole returns to the
Schwarzschild solution. In comparison with Schwarzschild black
hole, the CFM braneworld black hole will be either hotter or
colder depending upon whether $\gamma<3$ or $\gamma>3$.

We have calculated the scalar perturbation and the axial
gravitational perturbation around the brane black hole and
evaluated the energy absorptions of the scalar and gravitational
waves. Comparing with the 4D Schwarzschild black hole when
$\gamma=3$, we observed that the energy absorption of the
braneworld black hole decreases with the decrease of $\gamma$
starting from 3. While when $\gamma>3$ we found that the energy
absorption enhanced compared with that of the Schwarzschild black
hole. We restricted to the case $\gamma\leq 11/3$.  This result
holds the same for both the scalar wave and axial gravitational
wave outside the CFM hole. In both perturbations, the negative
potential well appeared, however the positive potential barrier
still dominated in determining the absorption rate, which has the
same effect as observed in the QNM study. The result on the
absorption spectrum implies that for $\gamma<3$, the black hole
emission will be enhanced with the decrease of $\gamma$, while for
$\gamma>3$, the emission of the black hole will be suppressed,
which is consistent with the behavior of the Hawking temperature
of the braneworld black hole. We conclude that the energy
absorption for the scalar and axial gravitational wave gives
signatures of the bulk effects in the brane-world black hole,
which differs from that of the 4D Schwarzschild black hole. We
expect that these signatures can be observed in the future
experiments, which could help us learn the properties of the extra
dimensions.

In the study of the axial gravitation perturbation, the deduction
of $\Psi_0^{(1)}=\tilde{\Psi}_0^{(1)}$ helps us a lot in doing the
calculation, however the simplification is obviously not hold for
polar perturbation. Thus it is of interest to generalize our
discussion in the future to study on the polar gravitation wave.

It needs to be emphasized that although several spherically
symmetric and static brane black hole solutions with contributions
from the bulk gravity have been found, none of these are obtained as
exact solutions of the full five-dimensional bulk field equations
\cite{s14}. The propagation of gravity into the bulk does not permit
the treatment of the brane gravitational field equations as a closed
system. This fact limits the consideration of propagating modes of
fields off the brane. Although the propagating modes in the bulk are
hard to be obtained at the present moment, the modes on the brane
are still interesting since they are the most phenomenologically
interesting effects which can be detected during experiments.
Furthermore, it was argued that the emission of particle modes on
the brane is dominant compared to that off the brane \cite{s10}.
Brane localized modes have been investigated to disclose the
information of extra-dimensions in different attempts
\cite{eabdalla,shen,s11,s12,park,park2}. Despite that we cannot
obtain the propagating modes in the bulk due to the lack of complete
bulk solutions owing to the conceptually complicated gravitational
field equations, our study of brane modes is still well-motivated
and interesting. Employing the Newman-Penrose formalism, we have
provided a test of the energy absorption spectrum of a new
braneworld black hole solution.

\begin{acknowledgments}
This work was partially supported by  NNSF of China, Ministry of
Education of China and Shanghai Science and Technology Commission.
We would like to acknowledge helpful discussions with J.Y. Shen.
\end{acknowledgments}

\end{document}